\documentclass[lettersize,journal]{IEEEtran}
\usepackage{amsmath,amsfonts}
\usepackage{algorithmic}
\usepackage{algorithm}
\usepackage{multirow}%后续添加
\usepackage[table]{xcolor}%后续添加
\usepackage{array}
\usepackage{textcomp}
\usepackage{stfloats}
\usepackage{subfigure}
\usepackage{url}
\usepackage{verbatim}
\usepackage{graphicx}
\usepackage{cite}
\usepackage{array}
\usepackage{colortbl}
\usepackage{makecell}
\usepackage[numbers]{natbib}
\usepackage{color}
\usepackage{pdfcomment}
\usepackage{enumitem}

\begin{document}

\title{Way to Build  Native AI-driven 6G Air Interface: Principles, Roadmap, and Outlook}

\author{
        \IEEEauthorblockN{Ping Zhang,~\IEEEmembership{Fellow Member,~IEEE,}        
                 Kai Niu,~\IEEEmembership{Member,~IEEE,}               
                 Yiming Liu,~\IEEEmembership{Member,~IEEE,} \\
                 Zijian Liang,~\IEEEmembership{Graduate Student Member,~IEEE,}
                 Nan Ma,~\IEEEmembership{Member,~IEEE,} \\
                 Xiaodong Xu,~\IEEEmembership{Senior Member,~IEEE,}  
                 Wenjun Xu,~\IEEEmembership{Senior Member,~IEEE,}  \\ 
                 Mengying Sun,~\IEEEmembership{Member,~IEEE,}  
                 Yinqiu Liu, 
                 Xiaoyun Wang,
                 and Ruichen Zhang,~\IEEEmembership{Member,~IEEE}}% <-this % stops a space
        \IEEEspecialpapernotice{(Invited Paper)}
        \thanks{This work was partly supported by the National Natural Science Foundation of China under Grants 62293480, 62293481, and 62471065. \textit{(Ping Zhang and Kai Niu contributed equally to this work and should be considered co-first authors. Co-corresponding authors: Kai Niu and Yiming Liu.)}}% <-this % stops a space
        \thanks{Ping Zhang, Yiming Liu, Nan Ma, Xiaodong Xu, Wenjun Xu, and Mengying Sun are with State Key Laboratory of Network and Switching Technology, Beijing University of Posts and Telecommunications, Beijing 100876, China. Ping Zhang, Nan Ma, Xiaodong Xu, and Wenjun Xu are also with Peng Cheng Laboratory, Shenzhen 518066, China (e-mail: pzhang@bupt.edu.cn, liuyiming@bupt.edu.cn, manan@bupt.edu.cn, xuxiaodong@bupt.edu.cn, wjxu@bupt.edu.cn, smy\_bupt@bupt.edu.cn).}% <-this % stops a space
        \thanks{Kai Niu and Zijian Liang are with Key Laboratory of Universal Wireless Communications, Ministry of Education, Beijing University of Posts and Telecommunications, Beijing 100876, China. Kai Niu is also with Peng Cheng Laboratory, Shenzhen 518066, China (e-mail: niukai@bupt.edu.cn).}% <-this % stops a space
        \thanks{Xiaoyun Wang is with China Mobile, Beijing 100032, China (e-mail: wangxiaoyun@chinamobile.com).}% <-this % stops a space
        \thanks{Yinqiu Liu and Ruichen Zhang are with the College of Computing and Data Science, Nanyang Technological University, Singapore (e-mail: yinqiu001@e.ntu.edu.sg, ruichen.zhang@ntu.edu.sg).}
}


\maketitle
\begin{abstract} 
Artificial intelligence (AI) is expected to serve as a foundational capability across the entire lifecycle of 6G networks, spanning design, deployment, and operation. This article proposes a native AI-driven air interface architecture built around two core characteristics: compression and adaptation. On one hand,  compression enables the system to understand and extract essential semantic information from the source data, focusing on task relevance rather than symbol-level accuracy. On the other hand, adaptation allows the air interface to dynamically transmit semantic information across diverse tasks, data types, and channel conditions, ensuring scalability and robustness. This article first introduces the native AI-driven air interface architecture, then discusses representative enabling methodologies, followed by a case study on semantic communication in 6G non-terrestrial networks. Finally, it presents a forward-looking discussion on the future of native AI in 6G, outlining key challenges and research opportunities.
\end{abstract}


% \fi
\IEEEpeerreviewmaketitle

\section{Introduction}

%\IEEEPARstart{T}{his} demo file is intended to serve as a ``starter file''
%for IEEE journal papers produced under \LaTeX\ using
%IEEEtran.cls version 1.8b and later.

The sixth generation (6G) of wireless networks is envisioned as a foundational transformation that extends far beyond incremental performance improvements. According to the International Telecommunication Union (ITU), 6G will support a new class of usage scenarios such as ubiquitous connectivity, integrated sensing and communication (ISAC), and artificial intelligence (AI) and communications \cite{series2023itu}. Unlike 5G, where AI is treated as an external enhancement, 6G positions AI as a native capability embedded throughout the entire network stack. In line with this vision, both academic and industry communities are exploring the development of native AI-driven 6G networks \cite{farhadi20256g, han2020artificial, zhang2025four, majumdar2025towards,sun2024s}. 

Although the integration of communications and AI is a natural and inevitable trend in 6G, achieving their cooperation through integrated design presents several critical challenges. Existing communication systems, which rely on layered architectures and modular signal processing chains, are not inherently compatible with the demands of native AI. Specifically, conventional model-based communication frameworks are not well-suited to fully leverage data-driven AI methodologies, limiting their effectiveness in dynamic environments. Moreover, the modular separation between components, such as source coding and channel coding, limits end-to-end optimization and decreases the overall system efficiency and adaptability required for complex wireless scenarios.

To address the above challenges, it becomes necessary to rethink the full-process communication systems. In this article, we propose a native AI-driven 6G air interface architecture. Unlike traditional systems where AI is applied as an optimization tool, this architecture integrates AI throughout the entire transmission process. In the proposed architecture, there are two core characteristics: compression and adaptation. First, to ensure efficiency, the native AI-driven 6G air interface must perform intelligent compression by understanding and extracting the essential semantics of the source data, rather than maintaining symbol-level accuracy. This allows for the transmission of concise, task-relevant information with reduced redundancy. Second, to support scalability, the native AI-driven 6G air interface must adaptively transmit semantic information across diverse tasks, data types, and channel conditions. This paradigm lays the foundation for a native AI-driven air interface capable of enabling intelligent, efficient, and context-aware communication.

The structure of this article is organized as follows. First, we present the native AI-driven 6G air interface architecture and highlight its core characteristics, focusing on compression and adaptation. Second, we introduce representative methodologies that enable the implementation of native AI within the 6G air interface. Next, we discuss a 6G non-terrestrial network (NTN) semantic communication system as a practical use case that illustrates the native AI-driven transmission in complex environments. Following that, we provide a forward-looking vision and outlook on the future of native AI, identifying key challenges and potential research directions. Finally, we conclude this article.

\section{Architecture and Core Principles of Native AI in 6G}

\subsection{Native AI-driven 6G Air Interface Architecture}

\begin{figure*}[!ht]
\centering
\includegraphics[width=0.85\textwidth]{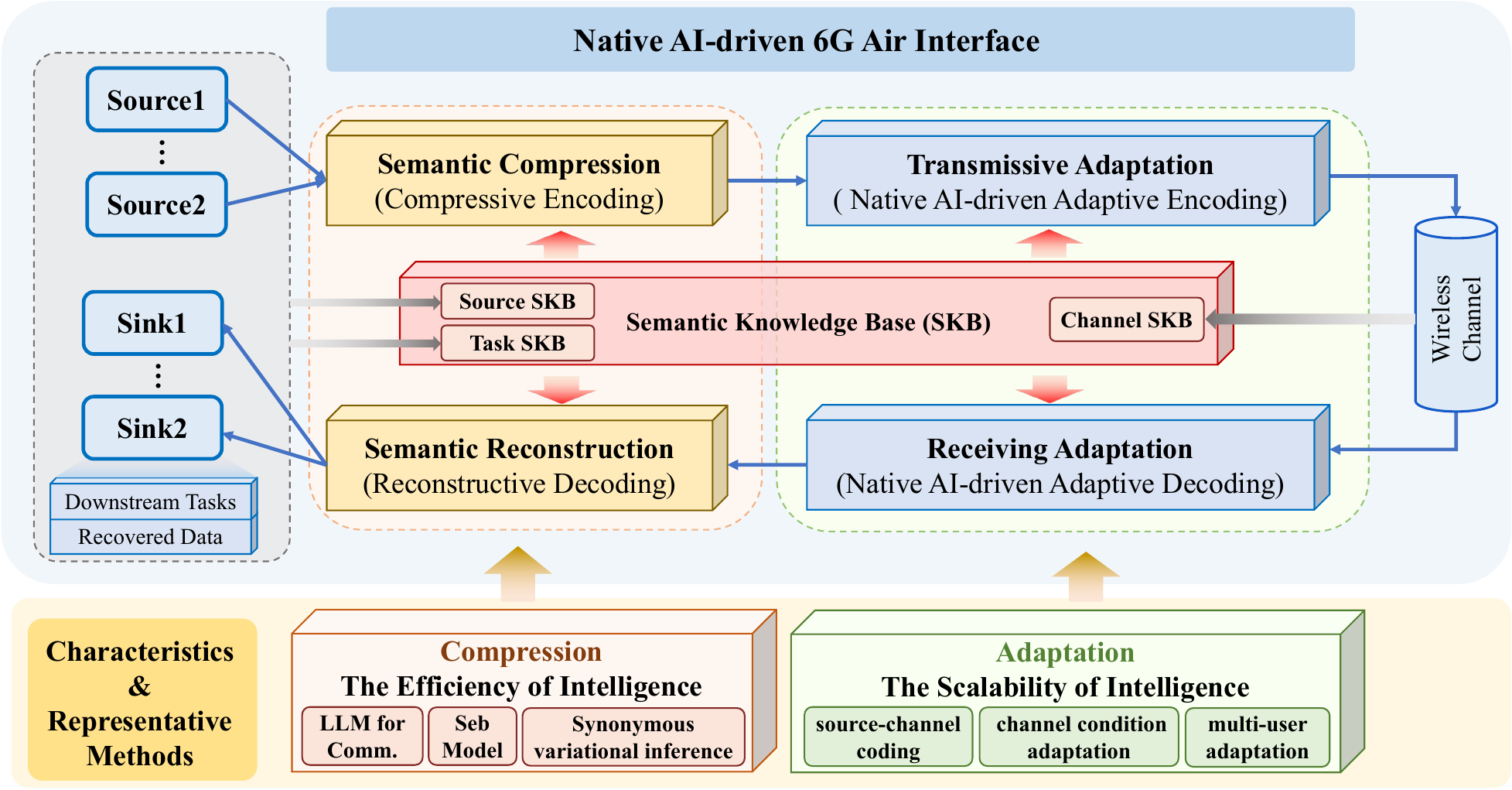}
\caption{The Native AI-driven Air Interface Architecture.}
\label{NativeAI_arc}
\end{figure*}

To support efficient information transmission across diverse scenarios and dynamic radio environments, the native AI-driven 6G air interface aims to move beyond conventional AI-assisted modular enhancements \cite{wang2025ai}. As illustrated in Fig. \ref{NativeAI_arc}, we propose a native AI-driven architecture that enables end-to-end learning of both the transmitter and receiver. 

At the core of the proposed architecture lies the integration of semantic compression and transmissive adaptation. Semantic compression aims to improve transmission efficiency by shifting the focus from raw bit delivery to the transmission of task-relevant meaning. This is achieved through compressive encoding, which extracts and compresses latent features that capture essential semantic content. Such encoding is typically realized using autoencoder-based neural networks jointly trained with entropy models, enabling compact representation and effective reconstruction within acceptable distortion levels. Compared with conventional coding schemes, this approach not only reduces redundancy but also aligns transmission objectives with the downstream tasks, such as inference or reconstruction, rather than simple data recovery. 
 
Transmissive adaptation, on the other hand, ensures that communication strategies remain robust in the face of varying channel conditions. By incorporating adaptive coding, modulation, mapping, etc., the system continuously aligns transmission parameters with the real-time state of the radio environment. This proactive adaptation enables the system to predict potential channel states and adjust its strategies accordingly, minimizing the impact of fading and interference. 
 
Within this architecture, the semantic knowledge base (SKB) plays a critical role in enabling efficient and context-aware communication between transmitter and receiver. SKBs function as structured and memorable knowledge networks that provide semantic knowledge descriptions of source data, channel environments, and task requirements \cite{zhang2024intellicise}. By organizing semantic knowledge into source, channel, and task knowledge bases, the SKB module allows the transmitter to jointly consider data semantics, channel conditions, and task contexts when performing adaptive source-channel coding, while the receiver leverages its local SKBs to reconstruct the transmitted meaning. This design transforms the mapping process from empirically defined symbol-to-bit encoding into a learning- and inference-driven mapping of symbols to semantic code streams, significantly improving efficiency and adaptability. To remain effective in dynamic environments, the SKB must support online learning and regional updates, ensuring adaptability to new scenarios. 
 
Ultimately, the native AI-driven 6G air interface transforms communication from a passive data delivery mechanism into an intelligent, context-aware system. By integrating semantic-level understanding with real-time adaptive transmission, it supports not only conventional human-centric services but also emerging machine-type and task-oriented applications. This architectural shift lays a critical foundation for realizing the 6G vision, enabling efficient, resilient, and intelligent wireless communication in the complex and heterogeneous environments expected in future networks.

\subsection{Core Characteristics}

\begin{figure*}[t]
\centering
\includegraphics[width=\textwidth]{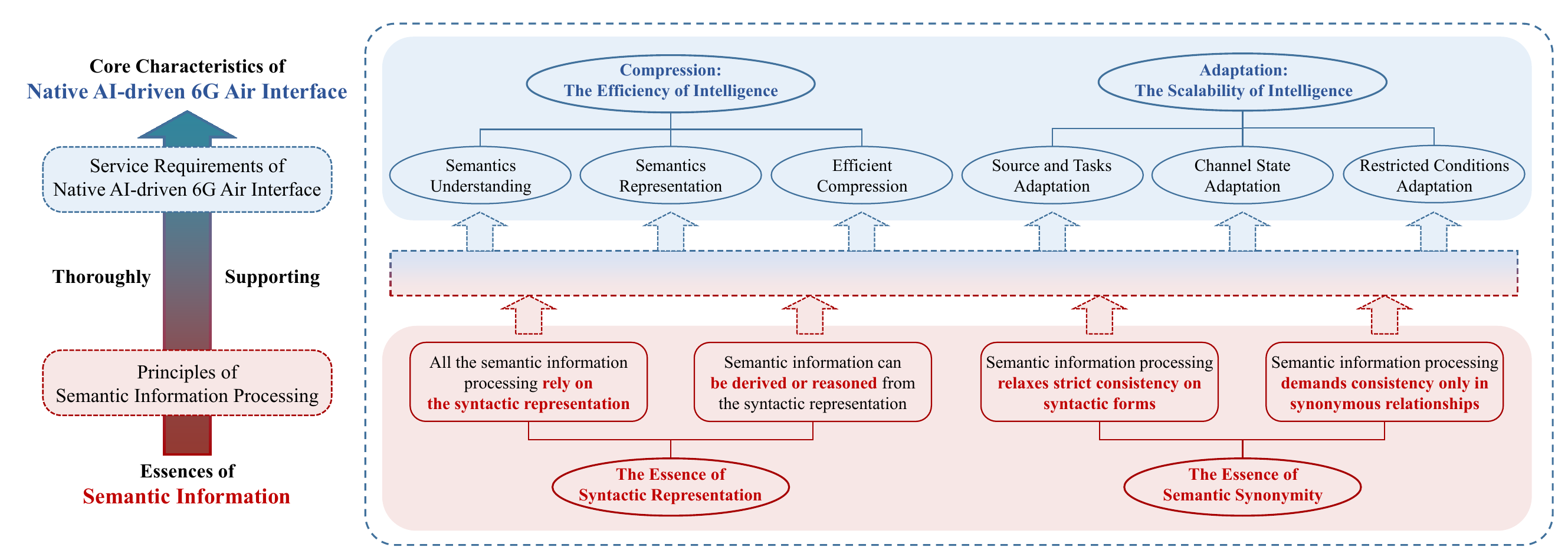}
\caption{The supporting relationship diagram of the essences of semantic information for the core characteristics of native AI-driven 6G air interface.}
\label{fig_SIT_supporting_NativeAI}
\end{figure*}

As communication technology evolves from a bit-centric paradigm to an AI-driven model, the core characteristics of the 6G air interface are expected to be founded on native AI. This native AI should be capable of deeply understanding the transmission contents and targets, and dynamically adapting to both source context features and the transmission environments, thereby enabling a more efficient and intelligent leap in information transmission capability compared to traditional wireless air interfaces. Specifically, the native AI-driven 6G air interface encompasses two core characteristics of intelligence from different perspectives:

\begin{itemize}
  \item \textbf{Compression: The Efficiency of Intelligence}. Recent AI research is revealing that compression is an essence of intelligence. In 2017, Tishby and Shwartz-Ziv \cite{shwartz2017opening} used information bottleneck theory to show that the compression phase in deep neural networks improves model generalization. In 2023, I. Sutskever \cite{sutskever2023observation} reveals the link between compression and intelligence when discussing unsupervised learning. Additionally, ongoing experiments and discussions \cite{Yu2023whitebox, deletang2024language, huang2024compression, li2025lossless} continue to confirm this viewpoint that higher intelligence yields stronger compression ability, whether through simple Transformer architecture or Transformer-based large-scale models. Based on this core characteristics of intelligence, the 6G air interface must use native AI to highly compress data: Unlike compression in traditional communications requiring symbol-level accuracy, the compression in the 6G air interface requires efficiently understanding the semantic characteristics of transmission services and requirements to represent the semantic information and achieve efficient compression, so that the required semantic information can be preserved as accurately as possible.
  
  \item \textbf{Adaptation: The Scalability of Intelligence}. This core characteristic was first proposed by P. Wang \cite{wang2007logic}, which distinguishes intelligent systems from the non-intelligent systems. It requires intelligent systems to be well-operated with insufficient knowledge and resources, under the restrictions of finite information-processing capability, real-time processing requirements, and no constraints on the knowledge and tasks that the system can accept. Based on these core characteristics, the 6G air interface is required to adapt to various data, tasks, and transmission environments, showing the scalability of the system. Specifically, scalability is reflected in generalizing unknown data and tasks \cite{sutskever2023observation, feng2023data} and dynamically adjusting to unfamiliar conditions \cite{joda2022internet, dai2023toward}. Thus, the 6G air interface is required to optimally transmit semantic information for any source data and task across diverse complex scenarios---adapting channel states and restricted conditions---which cannot be effectively achieved based on the bit-level reliable transmission of the traditional communication paradigm.
\end{itemize}

Corresponding to the evolution of communication technology, devices accessing the 6G air interface will be expected to shift from traditional user equipments to include various type of intelligent information processing devices and AI agents. At that time, intelligent processing will ease symbol-level accuracy demands and shift toward accurate semantic transmission.

By reviewing the fundamental theory of communication, it becomes clear that the semantic independence assumption of classical information theory \cite{shannon1948mathematical, weaver1949recent} becomes unsustainable for semantic information transmission, making semantic information theory essential for guiding the design and optimization of the 6G air interface to fully support these two core characteristics. However, classical views on semantic information—like those based on logical probability \cite{carnap1952outline, bar1953semantic, bao2011towards} or fuzzy sets \cite{de1972definition, de1974entropy, seising200960}—lack adequate theoretical support for these core characteristics and thus have inherent limitations. 

In 2024, a novel viewpoint on semantic information based on synonymity was introduced \cite{niu2024mathematical, niu2025mathematical}. By leveraging the ubiquitous phenomenon of synonymy in the natural world, it extends classical information theory into a comprehensive semantic information theoretical framework through the construction of synonymous relationships. It reveals two essences of the semantic processing, i.e., the essence of syntactic representation and the essence of the semantic synonymity, which establishes fundamental principles for semantic information handling. With these two essences thoroughly supporting the two core characteristics, as illustrated in Fig. \ref{fig_SIT_supporting_NativeAI}, this theoretical framework is expected to guide the design and optimization of native AI-driven 6G air interface:

\textbf{The essence of syntactic representation} implies that all the semantic information processing must rely on syntactic representation-whether it involves input data, intermediate results, or final outputs; however, the corresponding semantic information can be derived or reasoned from this syntactic representation. This essence supports the underlying semantic information carrying form of compression and adaptation in information processing-that is, both semantic representation and transmission waveform are expressed as the syntactic sequences. Moreover, this also provides foundational support for compatibility with existing communication system air interfaces.

\textbf{The essence of semantic synonymity} implies that a given semantic meaning can correspond to a set of synonymous syntactic sequences, interconnected through synonymous mappings. This implies that semantic information processing can relax strict consistency requirements on syntactic forms, demanding only consistency within synonymous relationships. This essence supports the fundamental mechanism of compression and adaptation in native AI-driven semantic information processing—that is, the synonymous relationships and syntactic carrying forms can be flexibly adjusted based on the understanding of the transmission content and targets, as well as dynamic adaptation to the source context features and the transmission environments.

\section{Representative Methods for Native AI-driven 6G Air Interface}

\subsection{Intelligence as Compression}
The compression characteristic captures the essence of native AI-driven communication, where the goal is not to transmit data with symbol-level precision but to convey the essential meaning with minimal redundancy.  In the 6G air interface, semantic compression and reconstruction can be added as new modules at the application layer, requiring communication systems to intelligently extract and represent the most task-relevant semantic information from the source to enable highly efficient transmission. Achieving such compression entails a deep understanding of both the transmitted content and the characteristics of the radio environment. This section explores representative approaches that address these challenges, including large-scale models for communication, semantic base (Seb) models, and synonymous variational inference. 

\subsubsection{Large-scale models for communication}

Large-scale models, such as large language models (LLMs) \cite{zhao2023survey} and vision-language models \cite{zhu2023minigpt}, offer a powerful tool for realizing intelligence as compression in native AI-driven 6G communication systems. These models possess strong capabilities in abstracting high-dimensional data into concise, task-relevant representations.  Rather than relying on handcrafted features or fixed encoding schemes, large-scale models can learn to extract semantic embeddings that preserve essential meaning while discarding redundant or irrelevant information. By focusing on transmitting only the semantically essential content required to complete a task, they significantly reduce bandwidth consumption while maintaining communication effectiveness. Beyond this, large-scale models demonstrate strong generalization across modalities, enabling unified compression of multimodal content, including text, images, and audio, thus supporting more complex intelligence tasks. In \cite{xiao2022imitation}, the authors propose a reasoning-based implicit semantic-aware communication network architecture in which multi-tier cloud data centers and edge servers collaborate to support the target users to directly learn a reasoning mechanism that automatically generates complex implicit semantic information based on limited cue information sent by the source users. In \cite{tian2024kg}, the authors investigate LLM agents for knowledge inference and question answering, which aims at improving answer accuracy for complex nested questions.

\subsubsection{Semantic base models}

Semantic base (Seb) models enable efficient and compact representation of diverse multimodal information by embedding semantic content within a structured semantic space \cite{zhang2022toward}. Acting as fundamental units, Sebs integrate background knowledge and communication intent with both explicit features and implicit, abstract attributes that capture semantics \cite{zhang2024advances}. Their granularity is adaptable to task requirements, with finer-grained Sebs supporting detailed reconstructions and coarser ones sufficing for classification tasks \cite{zhang2024intellicise}. In Seb-based communication systems,  the source message is first mapped into the semantic space through semantic extraction, obtaining corresponding hierarchical semantic features. These features are represented by Sebs within the knowledge bases, from which symbols are generated through semantic encoding. Notably, these features are composed of different Sebs with distinct levels of importance. The received symbols are interpreted by the receiver, aided by the knowledge bases, to recover the semantic features, which are then employed directly by the task-performing module to execute downstream tasks. This approach enables intelligent compression by balancing accuracy and efficiency, dynamically adjusting the granularity of Sebs according to context and communication goals. Zheng \textit{et al.} \cite{zheng2023semantic} proposed an explicit solution for Seb-based semantic communication, generating Sebs for images at both the transmitter and receiver, significantly enhancing transmission efficiency. The authors in \cite{wang2024semantic} proposed a new architecture of semantic communications based on explicit Sebs, where Sebs serve as the basic unit to describe semantic information. By establishing a structured and task-aware semantic representation, Seb models lay a solid foundation for future intelligent and adaptive communication systems that go beyond traditional bit-centric paradigms.

\subsubsection{Synonymous variational inference}

Synonymous variational inference \cite{liang2025synonymous} is a theoretical analysis method for semantic representation, offering a promising direction for designing semantic information compression methods guided by semantic information theory \cite{niu2024mathematical, niu2025mathematical}. Built upon the essences of syntactic representation and semantic synonymity, it enables both the extraction of a shared semantic representation for any sample within the synonymous set to which the source data belongs, and the generation of any sample within that set. This analytical approach reveals that perceptual similarity \cite{blau2018perception, blau2019rethinking} represents a typical form of synonymous representation, where different samples exhibiting perceptual similarity can be regarded as distinct syntactic instances within the same synonymous set. At the same time, the conclusions drawn from this analytical approach also provide theoretical support for the existence of different granularities and levels of Sebs. Although this analytical method currently offers only a preliminary theoretical exploration focused on image data, it is expected to be extended to various communication services involving other modalities and numerous downstream task-oriented scenarios, providing theoretical guidance and methodological support for the semantic compression requirements of native AI–driven 6G air interfaces.

\subsection{Intelligence as Adaptation}

The adaptation characteristic of the native AI-driven 6G air interface is primarily reflected in the transmission process, which is expected to be adapted to the transmission content and the channel environment to ensure optimal delivery of the required semantic information, showing the scalability of intelligence. From different perspectives, adaptation can be decomposed into several sub-problems, including adaptation in source-channel coding, adaptation to changing channel conditions, and adaptation in multi-user interference scenarios. This section presents representative methods corresponding to these three sub-problems.

\subsubsection{Adaptation on source-channel coding} 

Adaptation on source–channel coding refers to utilizing appropriate source and channel coding schemes that enable the transmission process to adapt to the source content or service requirements. Traditional communication systems use a separate source and channel coding architecture, ignoring semantic content and service demanding, treating the transmitted data as syntactic sequences. But under real conditions, where ideal assumptions fail, their end-to-end information transmission capability is inevitably limited \cite{vembu2002source, liang2024information}. In unstable channel conditions or poor transmission environments, it may even fail to effectively reconstruct the source data. Although some AI-for-wireless methods can be introduced, they typically use AI models to replace or optimize individual modules, yet this fundamental problem remains unresolved, as this module-by-module optimization approach still fails to overcome the limitations of traditional separate coding schemes. In contrast, the 6G air interface must adapt to both the transmission content and service requirements, making it necessary to adopt end-to-end joint source–channel coding (JSCC) methods.

Deep joint source–channel coding (DeepJSCC) is a class of early representative methods that employs a deep neural network to construct an autoencoder-like structure, jointly designing the source–channel encoder and decoder. By introducing channel noise at the bottleneck layer, the transmission process is made adaptive to the transmission content, guided by optimizing specific quality measures. This method was initially applied to data transmission across modalities such as text \cite{farsad2018deep}, images \cite{bourtsoulatze2019deep}, and speeches \cite{xie2021deep}, and has since served as a foundational approach guiding the design, optimization, and implementation of numerous subsequent techniques. However, this kind of approach has some significant limitations. While it overcomes the cliff effect \cite{bourtsoulatze2019deep} inherent in separate coding schemes, this straightforward approach was later shown to have limitations under high signal-to-noise ratio (SNR) and high-bandwidth conditions, resulting in end-to-end information transmission capacity that is lower than that of traditional methods.

Nonlinear transform source–channel coding (NTSCC) \cite{dai2022nonlinear} is another class of representative methods designed to overcome the limitations of DeepJSCC under high SNR and high-bandwidth conditions. It points out that as the embedding dimension increases, simpler feature vectors tend to waste more bandwidth during DeepJSCC transmission, while increasing SNR have a limited impact on the transmission qualities of these features vectors. To this end, NTSCC integrates a typical neural network-based image compression method—Nonlinear Transform Coding (NTC) \cite{balle2018variational, balle2020nonlinear}—with JSCC. First, it employs the entropy model to estimate the probability distribution and corresponding entropy value for each feature vector in the latent space. Then these feature vectors are used as input to the JSCC module, where the corresponding rate is adaptively allocated based on the feature vector’s entropy. This enables variable-rate transmission, allowing bandwidth to be allocated according to the importance of each feature vector and thereby achieving better end-to-end performance. This framework was later extended to modalities such as speech \cite{xiao2023wireless} and video \cite{wang2022wireless, yue2023learned}, in which the transmission performance was verified to surpass that of DeepJSCC approaches.

Furthermore, in evolving toward 6G air interfaces, compatibility with digital systems is a key direction for the adaptation on source-channel coding. This converts continuous encoder outputs into discrete constellation points for digital transmission, which extends the joint source-channel coding to joint coding-modulation \cite{zhang2025toward}. In practice, these methods are mainly divided into two categories: probabilistic modulation methods based on constellation point probability estimation \cite{bao2025sdac, bo2024joint}, and deterministic modulation methods based on scalar or vector quantization \cite{liu2024ofdm, fu2023vector}. These methods have been applied to DeepJSCC \cite{park2024joint} and NTSCC \cite{zhang2024analog}, even enabling them to adapt to varying channel conditions—an adaptability aspect discussed immediately afterwards. Such compatibility designs ensure that native AI-driven semantic communication methods can be effectively integrated into practical 6G air interfaces.

\subsubsection{Adaptation to changing channel conditions} 

Adaptation to changing channel conditions refers to the use of appropriate mechanisms that enable the transmission process to adapt to the varying channel conditions. The traditional air interface relies on processing mechanisms such as modulation and coding scheme (MCS) selection, beamforming/precoding, channel estimation, and channel state information (CSI) feedback, implementing system scheduling under strict protocol control to adapt to varying channel conditions \cite{takeda2020understanding}. Although such a mechanism can reduce the bit error rate to a certain extent and effectively support bit-level reliable transmission, its rigid protocol design struggles to adapt flexibly to rapidly varying and severely degraded channel conditions. To this end, the 6G air interface is required to be capable of adaptively adjusting the signal waveform in response to dynamic channel variations to maintain end-to-end transmission performance. It should be noted that this adaptation to changing channel conditions—since it requires adjusting the syntactic form for semantic information transmission—inevitably depends on the aforementioned adaptation on source–channel coding instead of completely decoupling from it.

A class of representative methods involves adjusting the transmission signal mechanism based on known channel information, including SNR adaptation, fading channel adaptation, and online learning techniques. SNR adaptation can typically be achieved by introducing attention mechanisms to adapt the transmission SNR—for example, by inserting an SNR attention module between layers of the DeepJSCC model \cite{xu2021wireless, xu2023deep} or by using a plug-and-play ModNet-like \cite{jia2022rate} module for adaptive adjustment \cite{yang2023witt, yang2024swinjscc}. Fading channel adaptation can not only adapt to fading channels by incorporating attention mechanisms \cite{wu2024deep, zhang2024scan, xie2024robust}, but also actively perform operations such as power allocation \cite{lin2023channel}, rate allocation, and antenna/stream mapping \cite{yao2023learned} based on the acquired channel state information (CSI) and supported by received signal processing or detection. Unlike the previous two methods that rely on offline training, online training techniques \cite{dai2023toward} can adjust model parameters—or directly modify the transmitted signal form—through brief learning in specific given fading channel scenarios, enabling real-time and rapid adaptation.

To effectively implement various signal waveform adaptation techniques, highly accurate CSI at both the transmitter and receiver is essential. To this end, a class of native AI-based CSI compression and feedback methods \cite{guo2022overview} have been proposed, including approaches based on deep neural network-based autoencoders \cite{wen2018deep, yang2019deep, cui2022transnet, yin2022deep}, vector-quantized variational autoencoders (VQ-VAE) \cite{shin2024vector, lee2024learning, shin2025entropy}, DeepJSCC \cite{xu2022deep, zheng2025semantic}, and so on. Some studies also explore the joint design of CSI feedback with channel estimation \cite{zhao2022joint, guo2024deep, feng2024deep}, beamforming/precoding \cite{guo2020deep, guo2024deep_multiuser, guo2025deep}, and related techniques to improve end-to-end transmission performance. In addition, recent research efforts have leveraged native AI to build a digital twin channel \cite{wang2024digital, wang2025radio} by building the relationship between channel information and the transmission environment. This approach maps radio channels and associated communication processes from the physical world into the digital domain in real time, which may support various applications and requirements in 6G air interface, such as channel prediction, channel semantic understanding, and efficient allocation of communication resources.

\subsubsection{Adaptation in multi-user scenarios}

Adaptation in multi-user scenarios refers to studying and designing mechanisms over 6G air interfaces using native AI-based approaches for multi-user resource scheduling and interference handling. Traditional air interfaces for multi-user communications rely mainly on orthogonal—or ideally orthogonal—multiple access \cite{shah2021survey} and duplexing \cite{yonis2012lte}, breaking multi-user communication into many point-to-point links. However, network information theory \cite{el2011network} shows that the overall channel capacity of multi-user systems can be further increased by exploiting characteristics of multi-user scenarios, such as multiple access, broadcasting, relaying, interference, and duplexing. Although traditional methods have been explored in these scenarios and some applied in practice \cite{ding2017survey, uyoata2021relaying, siddiqui2021interference, kolodziej2019band}, they still struggle to achieve optimal end-to-end information transmission performance (which can be modeled by ``feasibility'' \cite{zhang2023model, niu2024semantics} or called ``information conductivity'' \cite{liang2024information} of end-to-end communication systems) since these scenarios inherently fail to meet the source–channel separation theorem’s requirement for point-to-point links \cite{cover1999elements}. 

To this end, the native-AI-driven 6G air interface is expected to address these limitations by leveraging native AI to adapt to multi-user scenario characteristics, thereby enhancing the end-to-end information transmission capability of the entire multi-user communication scenario. For multiple access and broadcast scenarios, semantic orthogonality between models has been revealed, which enables representative methods to implement model-division multiple access \cite{zhang2023model} for multi-user end-to-end transmission based on different semantic communication models or processing directions \cite{zhang2023deepma, li2023non, mu2023exploiting, liang2024orthogonal}. For relay scenarios, edge intelligent relay nodes can enhance transmission links by introducing semantic knowledge bases \cite{luo2022autoencoder, lin2024semantic} or use semantic decoding and translation to deliver tailored information to users with different needs and service capabilities \cite{hu2024semantic}. For interference scenarios, semantic receivers can eliminate or suppress the semantic interference using a semantic knowledge base \cite{lin2024semantic} or applying native AI mechanisms like diffusion \cite{wu2025icdm}, which can also support multiple access designing \cite{ma2024semantic, zhang2024interference}. And for duplexing scenarios, semantic interference suppression in the semantic domain can be integrated into existing in-band full-duplex interference processing across spatial, analog, and digital domains, using generative mechanisms for reliable data reconstruction and enhancing overall end-to-end transmission capability \cite{niu2024semantics}. Furthermore, based on appropriate adjustments, the above native AI-driven multi-user semantic processing mechanisms can be applied to various distributed scenarios where signals from different users may not interfere but instead serve as side information, enabling link reconstruction via native AI to enhance communication \cite{mazandarani2024semantic, guo2024distributed, du2023ai}. These are expected to be widely applicable in multi-agent scenarios, establishing highly intelligent processing for future wireless networks.

\section{Experimental Platform: 6G NTN Semantic Communication System}

In this section, the establishment of the 6G NTN semantic communication system based on the digital video broadcasting-satellite-second generation (DVB-S2) protocol and the verification of semantic video transmission are presented to assess the transmission performance under the native-AI driven 6G air interface.

\subsection{6G NTN Semantic Communication System Setup}
\subsubsection{Hardware Components}
\begin{figure}[!t]
\centering
\includegraphics[width=0.45\textwidth]{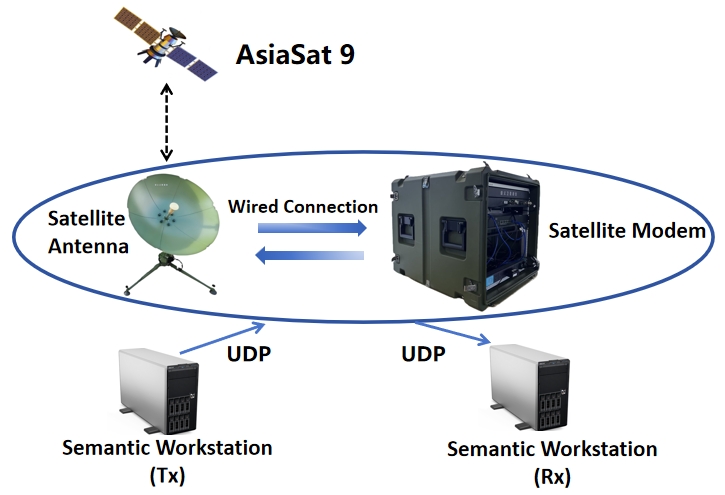}
\caption{The 6G NTN semantic communication system.}
\label{fig_6G_NTN}
\end{figure}

As shown in Fig. \ref{fig_6G_NTN}, the 6G NTN semantic communication system consists of the following hardware components:
\begin{itemize}
\item Two \textbf{semantic workstations}, each powered by an Intel Xeon Gold 6338 CPU and an Nvidia RTX 4090 GPU, one serving as the transmitting device and the other as the receiving device. The semantic workstation performs semantic extraction of the original video at the transmitter (Tx), and restores the distorted semantic information into the video at the receiver (Rx).
\item Two \textbf{satellite modems}, each establishes a wired connection with the semantic workstation and exchanges data via the user datagram protocol (UDP) protocol, followed by DVB-S2 modulation/demodulation of the transmitted data.
\item A \textbf{satellite antenna}, responsible for transmitting and receiving signals, enabling reliable space-to-ground data transmission. It features automatic tracking capability with azimuth/elevation/polarization adjustment to achieve precise satellite alignment. It integrates a BUC (Block Upconverter) for signal amplification, an LNB (Low-Noise Block) for downconversion.
\item The \textbf{AsiaSat 9}, functions as a data transponder, providing transparent transmission of data streams.
\end{itemize}

\begin{table*}[t]
\caption{Parameters of the 6G NTN semantic communication system}
\label{table_NTN_parameters}
\begin{center}
\begin{tabular}{|c|c|c|c|}
\hline
\textbf{Intermediate Frequency (IF)	}&1010 MHz&\textbf{Satellite-Ground Bandwidth}&1 MHz \\
\hline
\textbf{Block Up Converter (BUC)}& 13050 MHz&\textbf{Low Noise Block downconverter (LNB)}&11300 MHz\\
\hline
\textbf{Tx Rate}& 608 bps&\textbf{Rx Rate}&1296 bps\\
\hline
\textbf{Symbol Rate}& 800 kSps&\textbf{Tx Level}&-23.0 dBm\\
\hline
\textbf{Tx Semantic Workstation IP}& 192.168.10.100&\textbf{Tx Modem IP}&192.168.10.1\\
\hline
\textbf{Rx Semantic Workstation IP}& 192.168.11.100&\textbf{Rx Modem IP}&192.168.11.253\\
\hline
\textbf{Antenna IP}& 192.168.100.7&\textbf{SNR}&9.3-10.0dB\\
\hline
\end{tabular}
\label{tab1}
\end{center}
\end{table*}

\subsubsection{System Transmission Process}
At the Tx, the semantic workstation performs semantic extraction and joint source and channel encoding of the source data, then transmits the processed stream via UDP protocol to the modulator. After DVB-S2 modulation, the signal is power-amplified and uplinked through the satellite antenna. AsiaSat 9 operates as a transparent transponder, relaying the signal back to Earth where it is received by the ground station antenna. The received signal is routed to the demodulator for DVB-S2 demodulation, then forwarded via UDP to the receiving semantic workstation. Through semantic decoding and reconstruction, the original source data can be recovered. The detailed parameters of the 6G NTN semantic communication system are shown in TABLE \ref{table_NTN_parameters}.

\begin{figure}[t]
\centering
\includegraphics[width=0.45\textwidth]{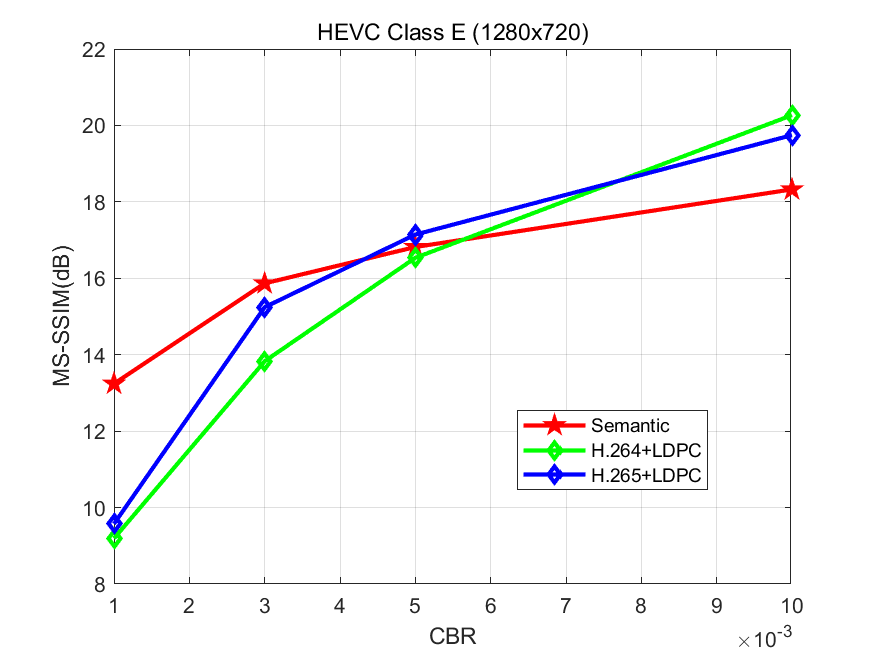}
\caption{MS-SSIM performance comparison versus CBR over the real GEO channel.}
\label{fig_MSSSIM_CBR_GEO_channel}
\end{figure}

\begin{figure*}[hbtp]
\centering
        \begin{minipage}{0.37\linewidth}
		\vspace{3pt}
		\centerline{\includegraphics[width=\textwidth]{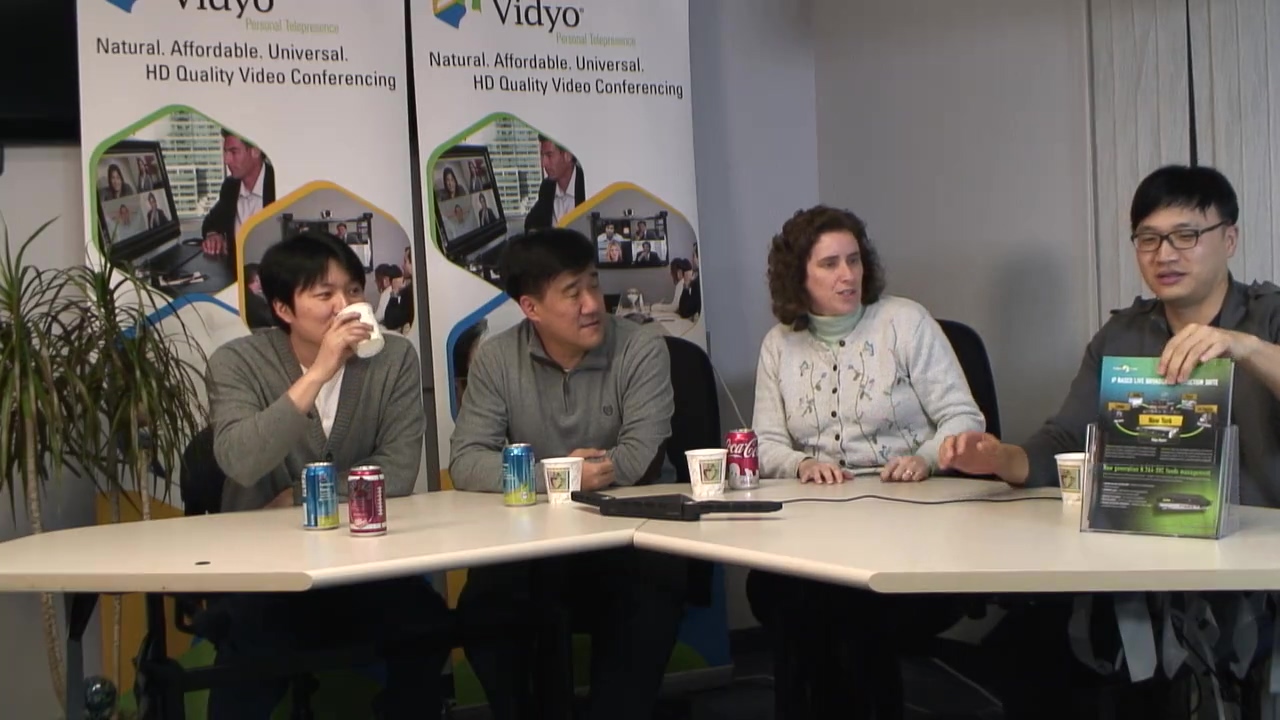}}
        \centerline{(a) Original Frame}
	\end{minipage}
	\begin{minipage}{0.37\linewidth}
		\vspace{3pt}
        \centerline{\includegraphics[width=\textwidth]{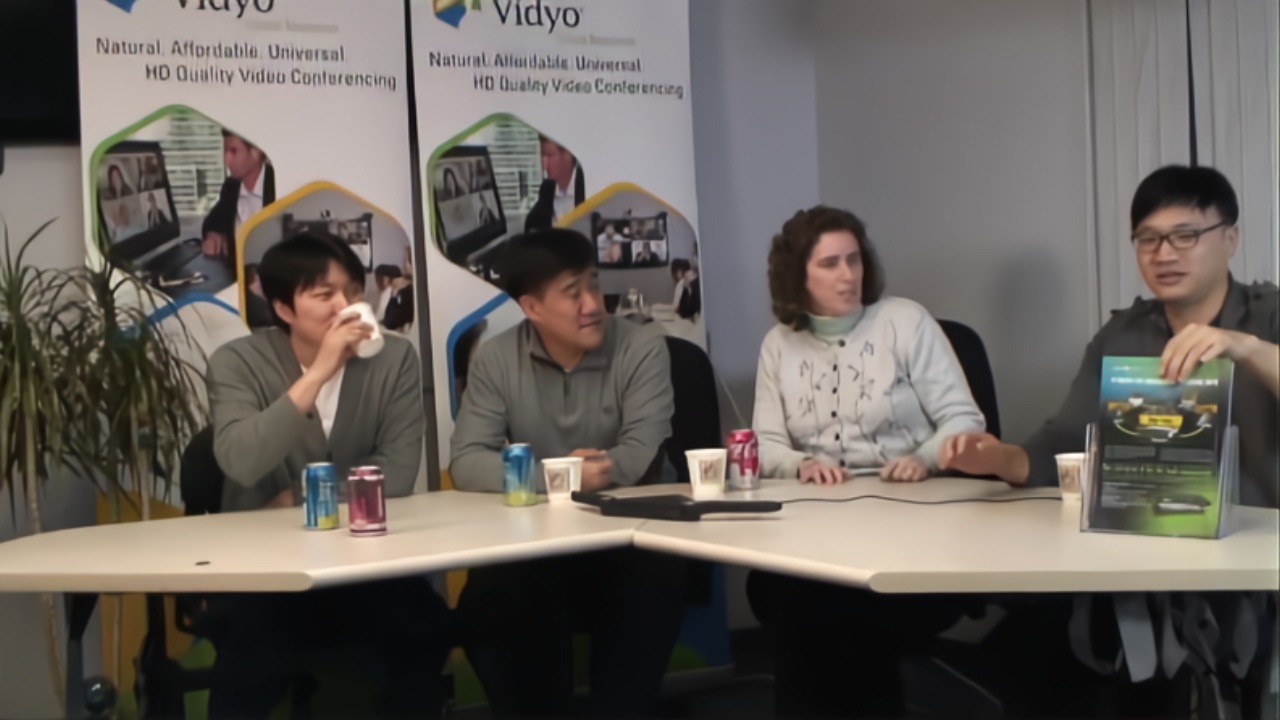}}
        \centerline{(b) Semantic Video Transmission}
	\end{minipage}
    
    \begin{minipage}{0.37\linewidth}
		\vspace{3pt}
        \centerline{\includegraphics[width=\textwidth]{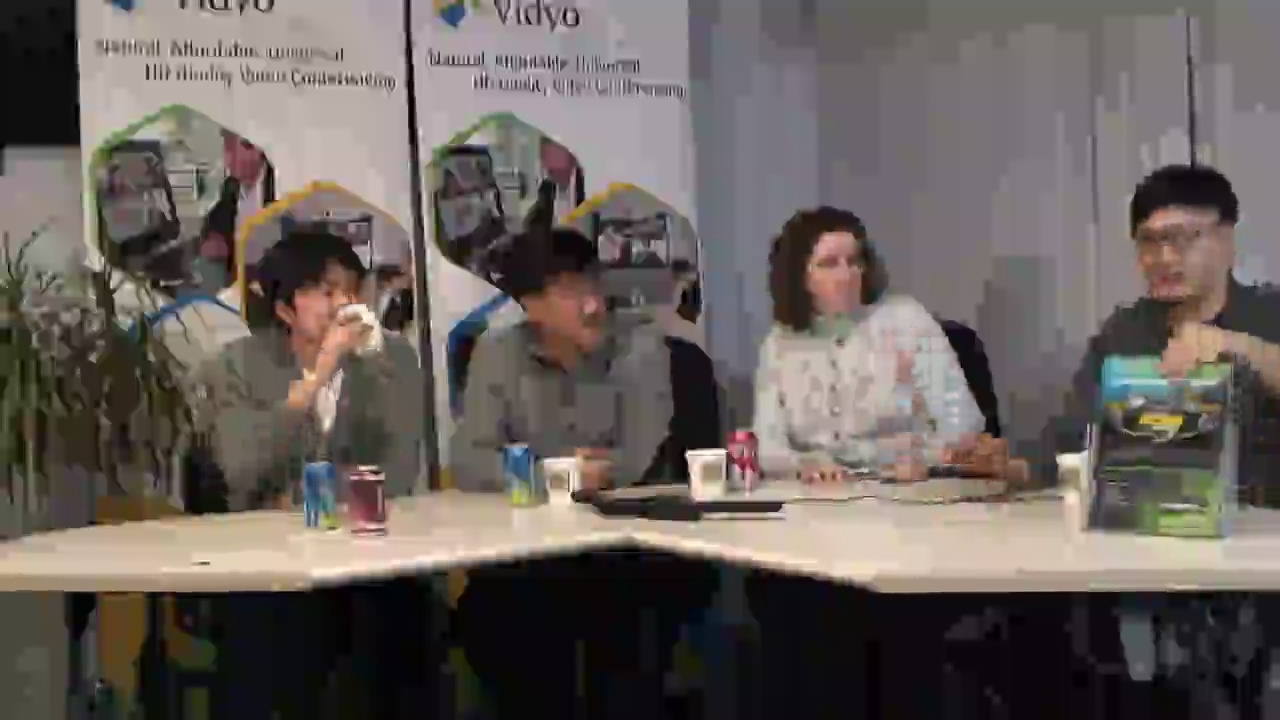}}
        \centerline{(c) H.264 + LDPC}
	\end{minipage}
	\begin{minipage}{0.37\linewidth}
		\vspace{3pt}
		\centerline{\includegraphics[width=\textwidth]{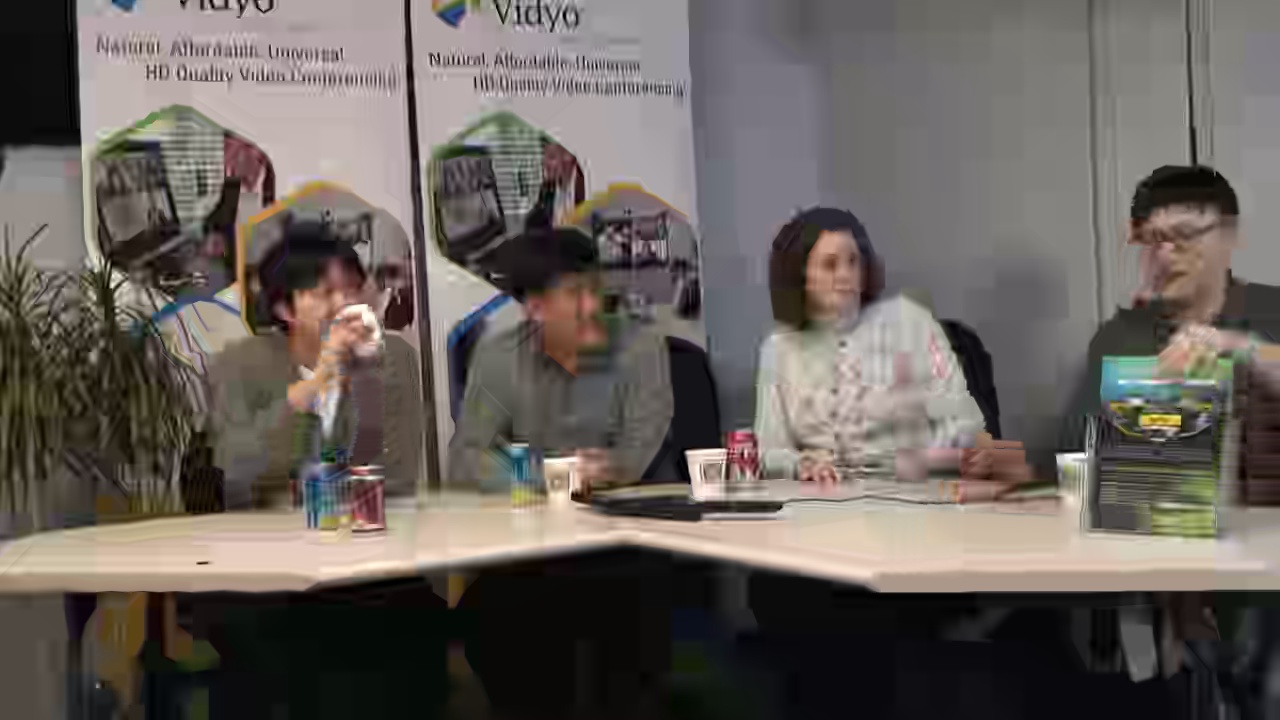}}
        \centerline{(d) H.265 + LDPC}
	\end{minipage}
\caption{Examples of visual comparisons at CBR = 0.001.}
\label{fig_visualization_CBR}
\end{figure*}

\subsection{Experimental Setup}
In the experiments, the satellite-to-ground transmission validation for video semantic communication systems is conducted.
Experiments were conducted utilizing AsiaSat 9 (122°E) as the communication satellite. The channel was a geostationary Earth orbit (GEO) satellite link with dynamic SNR ranging from about 9.3dB to 10.0dB. The tested video is selected from the standard HEVC test datasets \cite{bossen2010common}.
The video semantic coding employs the adaptive frame interpolation video semantic communication scheme \cite{qi2024wafi}. The channel bandwidth ratio (CBR) \cite{kurka2020deepjscc} which is defined as the ratio of channel coding rate to source coding rate is used to measure the cost of communication resources, which is formulated as ${n_c}/{n_s}$, in which $n_c$ represents the number of channel symbol, and $n_s$ represents the number of source symbol.

The performance is tested using multi-scale structure similarity (MS-SSIM)\cite{wang2003multiscale} as an evaluation metric. MS-SSIM takes into account the perception of structural information by the human visual system, enabling it to better capture details and texture information in images and evaluate the perceptual quality of the recovered frames. Since the values of MS-SSIM mostly exceed 0.9, the formula $-10log_{10}(1-d)$ can be utilized to convert them into dB for a more intuitive representation of these results, where $d$ is the value of MS-SSIM \cite{balle2018variational}.

The traditional compared schemes include source coding using advanced video coding (AVC/H.264) \cite{wiegand2003overview} and high efficiency video coding (HEVC/H.265) \cite{sullivan2012overview}, channel coding using (720,1440) low-density parity-check (LDPC) coding \cite{gallager2003low} with a coding rate of 1/2. And the modulation scheme uses BPSK.

\subsection{Experimental Results and Performance Analysis}
\subsubsection{Reconstruction Quality at Different CBRs}
As shown in Fig. \ref{fig_MSSSIM_CBR_GEO_channel}, the semantic video transmission scheme has significant advantages over the traditional H.264 + LDPC scheme under extremely low transmission volume. To achieve the same effect (CBR = 0.001), the transmission efficiency is improved by about 3 times.

In the visualization results of Fig. \ref{fig_visualization_CBR}, it can be observed that the semantic video transmission scheme outperforms traditional methods in both overall and local performance at extremely low CBR, which demonstrates the compression capability and the adaptability of native AI under bandwidth-constrained conditions.

\subsubsection{Reconstruction Quality at Different SNRs}
\begin{figure}[t]
\centering
\includegraphics[width=0.45\textwidth]{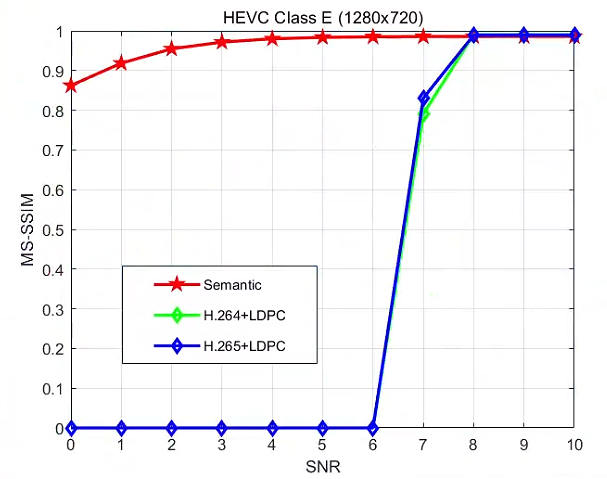}
\caption{MS-SSIM performance comparison versus SNR over the GEO channel.}
\label{fig_MSSSIM_SNR_GEO_channel}
\end{figure}

The satellite SNR is controlled by means of secondary noise addition in cascaded channels.
As shown in Fig. \ref{fig_MSSSIM_SNR_GEO_channel}, under high SNRs (SNR $\ge$ 8dB), the semantic video transmission scheme achieves the same performance as the traditional H.264 and H.265 schemes. As the SNR decreases, starting from 7dB, the traditional schemes, affected by the cliff effect, begin to drop sharply, while the semantic video transmission scheme exhibits a steady decline. Even at 0dB, it still achieves an MS-SSIM of more than 0.85. 

\begin{figure*}[hbtp]
\centering
        \begin{minipage}{0.37\linewidth}
		\vspace{3pt}
		\centerline{\includegraphics[width=\textwidth]{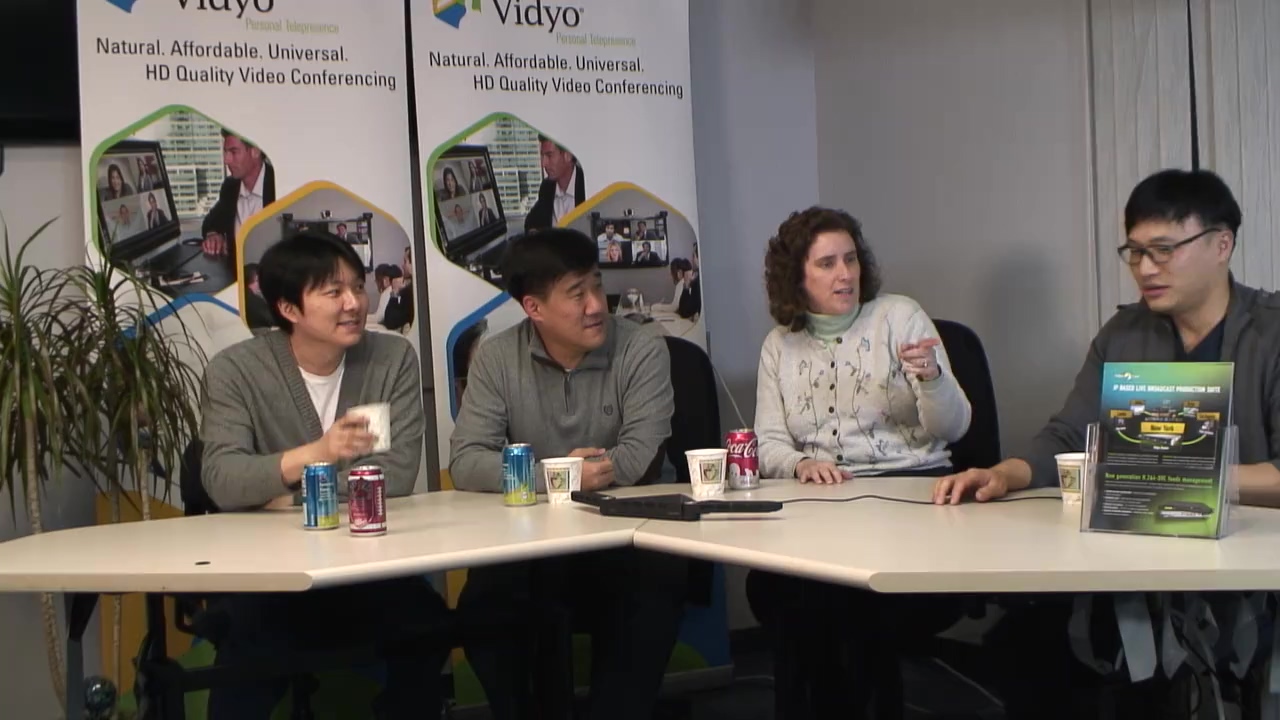}}
        \centerline{(a) Original Frame}
	\end{minipage}
    \begin{minipage}{0.37\linewidth}
		\vspace{3pt}
		\centerline{\includegraphics[width=\textwidth]{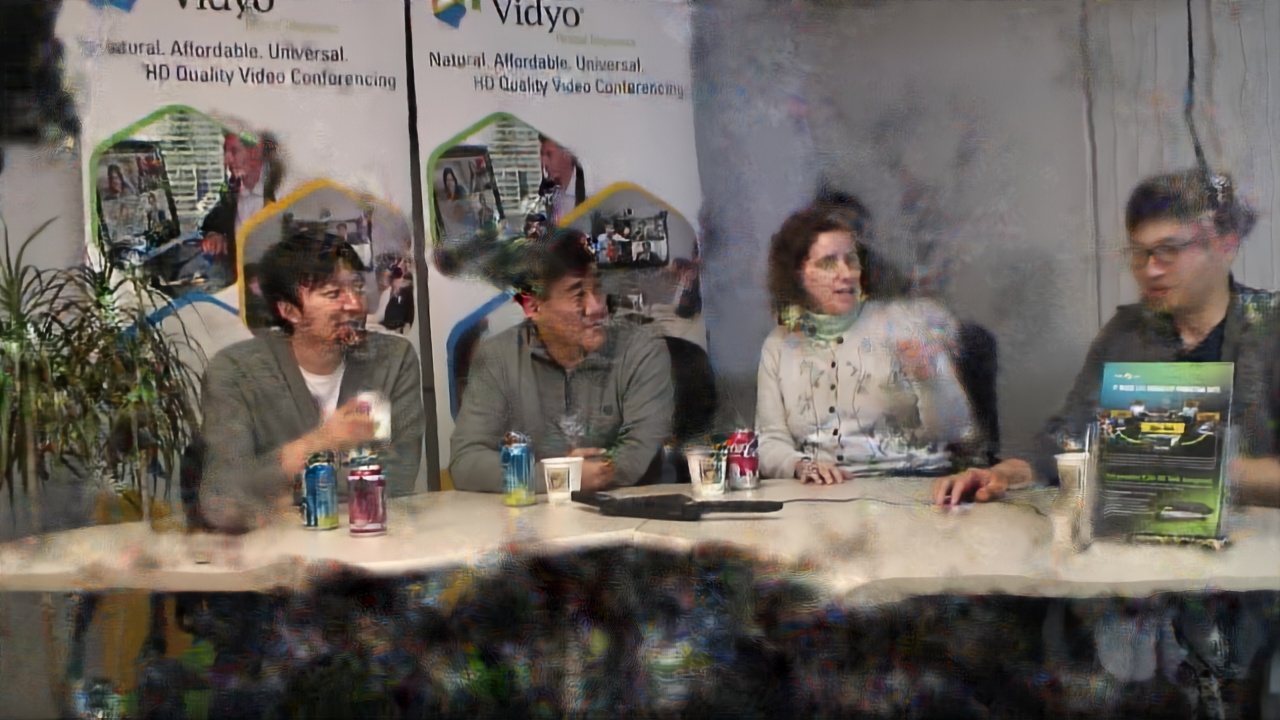}}
        \centerline{(b) -5dB Semantic video transmission}
	\end{minipage}
    
    \begin{minipage}{0.37\linewidth}
		\vspace{3pt}
        \centerline{\includegraphics[width=\textwidth]{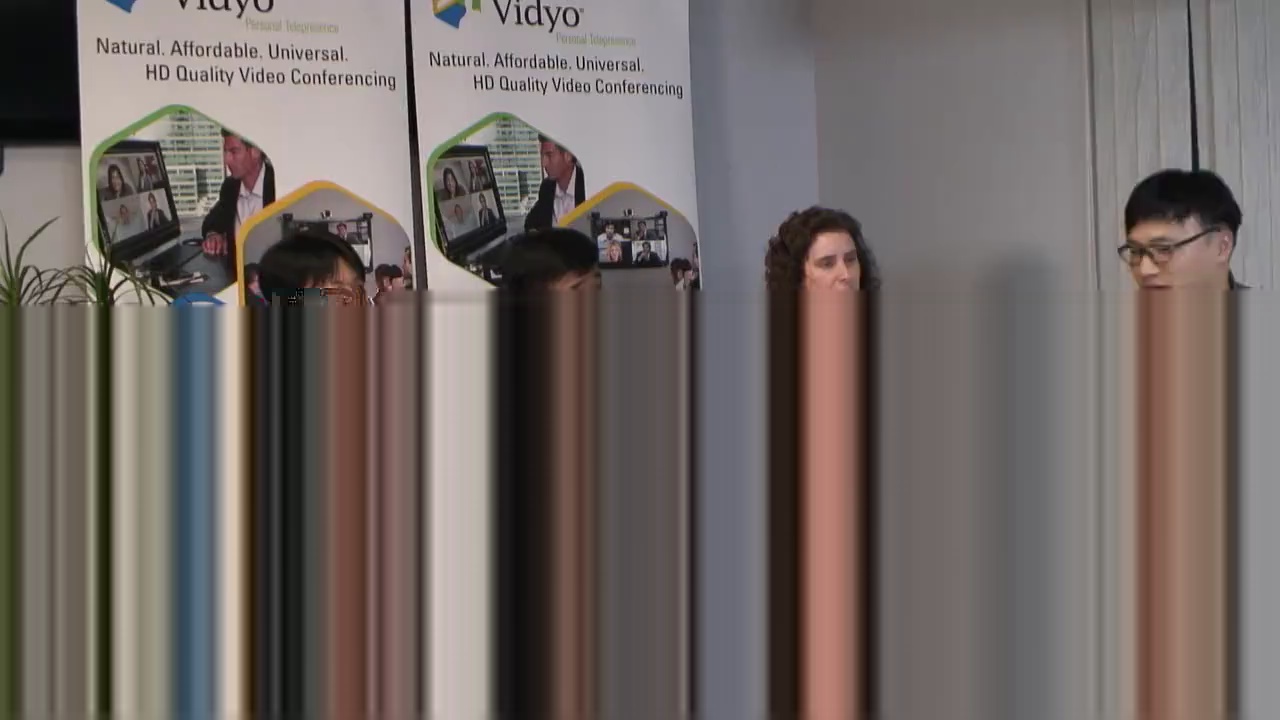}}
        \centerline{(c) 7dB H.264 + LDPC}
	\end{minipage}
    \begin{minipage}{0.37\linewidth}
		\vspace{3pt}
        \centerline{\includegraphics[width=\textwidth]{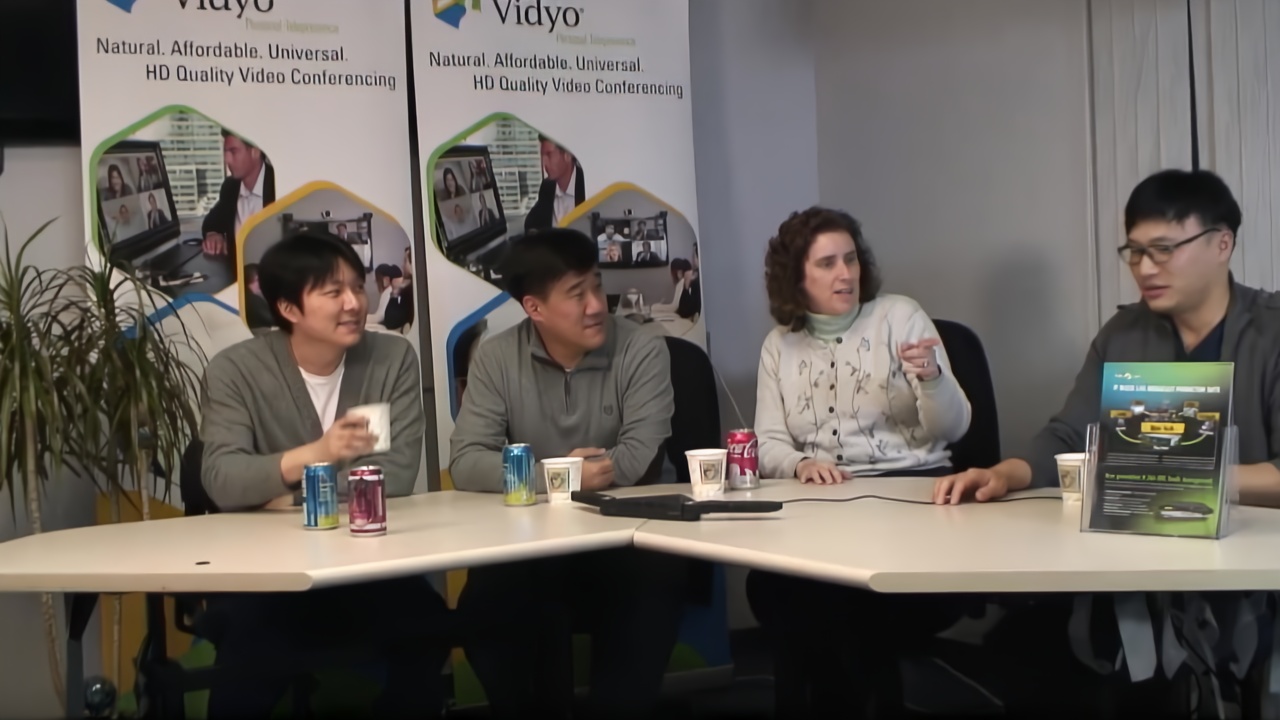}}
        \centerline{(d) 7dB Semantic video transmission}
	\end{minipage}
	
\caption{Examples of visual comparisons at different SNRs.}
\label{fig_visualization_SNR}
\end{figure*}

In the visualization results of Fig. \ref{fig_visualization_SNR}, it can be observed that the restored video of the traditional H.264 + LDPC scheme has serious distortion problems. Obvious vertical stripes and color distortion appear in the lower part of the picture, which seriously affects the clarity and viewing experience of the video. The semantic video transmission scheme has a very clear restoration effect at 7dB. Even at an equivalent -5dB, there is only partial regional blurriness, and the video content can still be roughly understood. This demonstrates native AI’s adaptability under varying and limited channel conditions.

\section{Vision and outlook for Native AI in 6G}

\subsection{Native AI as the foundation of future air interface design}

\subsubsection{System-level Performance Optimization}
 In 6G scenarios, many applications such as ultra-reliable low-latency communication (URLLC), autonomous driving, and industrial automation impose stringent latency constraints, where AI-driven information processing and signal transmission must be completed within milliseconds. Achieving such low latency often necessitates simplifying AI models through quantization or pruning, which can compromise their ability to model complex patterns or adapt to rapidly changing channel environments. This trade-off is further complicated by practical hardware constraints. AI models deployed at the network edge or on user equipment (UE) must operate within strict limits on computation, memory, and energy. While recent advances in AI-specific chipsets, such as inference accelerators using 8-bit or 4-bit arithmetic, provide significant improvements in energy efficiency and throughput, they remain limited in their ability to support high-precision models. To overcome these challenges, several strategies have been proposed. One is the adoption of lightweight neural network architectures, which reduce complexity without severely compromising performance. Techniques such as model pruning remove redundant parameters, while knowledge distillation transfers knowledge from large models to compact ones better suited for edge deployment. Quantization-aware training further enables low-precision models to maintain accuracy comparable to their full-precision counterparts.  

\subsubsection{New design paradigm and Implementation}

While native AI is envisioned as a foundational capability in 6G air interface, its practical implementation poses significant challenges. Current communication systems typically rely on separately designed source and channel coding modules, which operate independently and are optimized in isolation. Integrating AI into this framework requires a substantial redesign of the entire communication stack, increasing system complexity and demanding additional computational and energy resources. Moreover, to enable coexistence with existing networks, AI-native air interfaces must coordinate with conventional network elements such as scheduling, resource allocation, and interference management. Seamless fallback mechanisms are essential to allow devices to dynamically switch between AI-native and traditional communication modes based on real-time conditions. Furthermore, maintaining the accuracy and relevance of AI models in 6G air interface necessitates continuous data collection, secure over-the-air model updates, robust version control, and on-demand retraining. However, frequent model updates can strain device storage, increase latency, and conflict with energy constraints, especially in power-limited edge devices. Large-scale AI operations can significantly increase the overall energy consumption of the network. To address this, joint optimization of communication and computation resources, guided by semantic importance,  will be crucial in achieving both performance and sustainability goals.

\subsection{Open research challenges: generalization, trustworthiness, complexity}
\subsubsection{Generalization} Generalization remains a central challenge for the native AI-driven 6G air interface. While existing techniques—such as attention mechanisms and resource allocation based on feedback CSI—can enable reliable channel adaptation, the 6G air interface must be capable of adapting to diverse, highly dynamic, and often unpredictable real-world conditions, including new environments, devices, mobility patterns, and traffic types. Current AI models typically perform well on scenarios resembling their training data but struggle under out-of-distribution conditions. Achieving robust generalization across heterogeneous deployment scenarios demands innovations in training paradigms—such as meta-learning, domain adaptation, and self-supervised learning—alongside large-scale, high-fidelity simulation environments that better capture the variability encountered in practical networks.
\subsubsection{Trustworthiness} Trustworthiness in native AI-driven 6G air interfaces involves safety, security, privacy, resilience, and reliability—crucial for dependable communication. Unlike traditional modular systems, AI components act as opaque black boxes, making it difficult to guarantee safe and trustworthy behavior under abnormal or adversarial conditions, even when protective mechanisms are in place. End-to-end AI models that map application data to channel signals also risk privacy data leakage and attacks like adversarial perturbations. In addition, cryptographic integration is also challenging: traditional encryption can disrupt model training, while learned encryption lacks formal guarantees. Moreover, ensuring resilience and long-term reliability in dynamic environments demands a fundamental rethink of secure AI design, privacy-preserving learning, and hybrid architectures combining cryptographic rigor with deep learning flexibility. Therefore, establishing corresponding theories, mechanisms, and methods that leverage native AI capabilities to ensure the trustworthiness of 6G air interfaces remains a critical challenge requiring focused attention in future research.
\subsubsection{Complexity} Complexity is inherently tied to the integration of AI into the 6G air interface, both in algorithmic and system-level terms. AI-driven designs often introduce considerable computational and storage demands, especially when relying on large-scale models or real-time inference. These demands challenge the resource constraints of edge devices and distributed radio access networks. Furthermore, managing the interaction between data-driven AI components and traditional model-based modules adds layers of architectural and operational complexity. Coordinating learning across decentralized nodes, ensuring synchronization, and handling heterogeneous hardware capabilities make the end-to-end system design significantly more difficult. Addressing these issues requires scalable, efficient, and modular solutions—such as lightweight neural architectures, edge–cloud co-design, and federated learning—that are expected to manage complexity without sacrificing performance under native intelligent design and optimization control.

\subsection{Steps toward standardization and large-scale deployment}
\subsubsection{Unified Architecture and Compatibility Design}
To support the standardization of SemCom, a unified end-to-end foundational framework should be established, with clearly defined functional modules for semantic encoding, decoding, extraction, inference, and feedback. The architecture should also ensure compatibility with existing mobile communication systems to enable seamless deployment of semantic enhancements. For practical deployment, key functional modules within the access and core networks must be designed, along with the management of associated protocols and implementation procedures. Moreover, this unified framework provides essential infrastructure for diverse application scenarios and facilitates the development of AI-native 6G systems. To support gradual adoption, standardization is progressively expanding from the control plane (e.g., CSI compression and feedback) to the data and user planes.

\subsubsection{Unified Evaluation Metrics and Validation Platform}
To support the standardization of SemCom, a quantifiable, comparable, and reproducible evaluation framework must be established. This involves defining standardized and integrated semantic performance metrics, including task completion rate, semantic fidelity, intent alignment accuracy, as well as conventional communication metrics. Standardized datasets and representative task scenarios should be developed to support model training and benchmarking across diverse use cases. In addition, reproducible simulation platforms and end-to-end testbeds should be designed to emulate realistic communication environments, incorporating factors such as channel fading, interference, and mobility. This standardized evaluation infrastructure is essential for technology validation, performance comparison, and compliance assessment, ultimately accelerating the translation of research innovations into deployable industrial solutions.

\section{Conclusion }

Native AI redefines the foundation of future 6G air interfaces by embedding intelligence into the communication process. This article has presented a native AI-driven 6G air interface architecture built around two core characteristics: compression and adaptation. Compression enables the extraction and transmission of task-relevant semantic information, significantly reducing redundancy and enhancing communication efficiency. Adaptation allows the interface to dynamically respond to diverse channel conditions, tasks, and user requirements, ensuring robustness and scalability. We first outlined the overall architecture, then introduced key enabling technologies and presented a case study on semantic communication in non-terrestrial networks. Finally, we discussed several open challenges and future research directions related to native AI in 6G. 

\ifCLASSOPTIONcaptionsoff
  \newpage
\fi
\bibliographystyle{IEEEtran}
\bibliography{ref}

\begin{thebibliography}{100}
\providecommand{\url}[1]{#1}
\csname url@samestyle\endcsname
\providecommand{\newblock}{\relax}
\providecommand{\bibinfo}[2]{#2}
\providecommand{\BIBentrySTDinterwordspacing}{\spaceskip=0pt\relax}
\providecommand{\BIBentryALTinterwordstretchfactor}{4}
\providecommand{\BIBentryALTinterwordspacing}{\spaceskip=\fontdimen2\font plus
\BIBentryALTinterwordstretchfactor\fontdimen3\font minus
  \fontdimen4\font\relax}
\providecommand{\BIBforeignlanguage}[2]{{%
\expandafter\ifx\csname l@#1\endcsname\relax
\typeout{** WARNING: IEEEtran.bst: No hyphenation pattern has been}%
\typeout{** loaded for the language `#1'. Using the pattern for}%
\typeout{** the default language instead.}%
\else
\language=\csname l@#1\endcsname
\fi
#2}}
\providecommand{\BIBdecl}{\relax}
\BIBdecl

\bibitem{series2023itu}
\BIBentryALTinterwordspacing
ITU-R, ``Framework and overall objectives of the future development of imt for
  2030 and beyond.'' [Online]. Available: \url{https://www.itu.int/rec/R-REC-
  M.2160-0-202311-I/en}
\BIBentrySTDinterwordspacing

\bibitem{farhadi20256g}
H.~Farhadi, B.~Banerjee, R.~Berkvens, N.~N. Bhat, E.~Bodji, D.~Dampahalage,
  E.~Eldeeb, J.~Famaey, G.~P. Fettweis, J.~Jeong \emph{et~al.}, ``{6G}
  {AI}-driven air interface—{Hexa-X-II} view,'' \emph{IEEE Communications
  Magazine}, 2025.

\bibitem{han2020artificial}
S.~Han, T.~Xie, I.~Chih-Lin, L.~Chai, Z.~Liu, Y.~Yuan, and C.~Cui,
  ``Artificial-intelligence-enabled air interface for {6G}: Solutions,
  challenges, and standardization impacts,'' \emph{IEEE Communications
  Magazine}, vol.~58, no.~10, pp. 73--79, 2020.

\bibitem{zhang2025four}
J.~Zhang, Y.~Cai, L.~Yu, Z.~Zhang, Y.~Zhang, J.~Wang, T.~Jiang, L.~Xia, and
  P.~Zhang, ``Four steps toward {6G} {AI}-enabled air interface: Wireless
  environmental information sensing, feature, semantics, and knowledge,''
  \emph{IEEE Communications Magazine}, vol.~63, no.~8, pp. 56--62, 2025.

\bibitem{majumdar2025towards}
S.~Majumdar, Q.~Wei, S.~Schwarzmann, R.~Trivisonno, and G.~Carle, ``Towards
  {AI}-native {6G} systems: Standards enablers for {6G} network automation,''
  \emph{IEEE Communications Standards Magazine}, 2025.

\bibitem{sun2024s}
Y.~Sun, L.~Zhang, L.~Guo, J.~Li, D.~Niyato, and Y.~Fang, ``S-{RAN}:
  Semantic-aware radio access networks,'' \emph{IEEE Communications Magazine},
  2024.

\bibitem{wang2025ai}
X.~Wang, S.~Han, Z.~Liu, Q.~Wang, J.~Wang \emph{et~al.}, ``{AI}-driven {6G} air
  interface: Technical usage scenarios and balanced design methodology,''
  \emph{SCIENTIA SINICA Informationis}, vol.~55, no.~6, pp. 1522--1533, 2025.

\bibitem{zhang2024intellicise}
P.~Zhang, W.~Xu, Y.~Liu, X.~Qin, K.~Niu, S.~Cui, G.~Shi, Z.~Qin, X.~Xu, F.~Wang
  \emph{et~al.}, ``Intellicise wireless networks from semantic communications:
  A survey, research issues, and challenges,'' \emph{IEEE Communications
  Surveys \& Tutorials}, 2024.

\bibitem{shwartz2017opening}
R.~Shwartz-Ziv and N.~Tishby, ``Opening the black box of deep neural networks
  via information,'' \emph{arXiv preprint arXiv:1703.00810}, 2017.

\bibitem{sutskever2023observation}
I.~Sutskever, ``An observation on generalization,'' in \emph{Workshop on Large
  Language Models and Transformers}, 2023.

\bibitem{Yu2023whitebox}
\BIBentryALTinterwordspacing
Y.~Yu, S.~Buchanan, D.~Pai, T.~Chu, Z.~Wu, S.~Tong, H.~Bai, Y.~Zhai, B.~D.
  Haeffele, and Y.~Ma, ``White-box transformers via sparse rate reduction:
  Compression is all there is?'' \emph{Journal of Machine Learning Research},
  vol.~25, no. 300, pp. 1--128, 2024. [Online]. Available:
  \url{http://jmlr.org/papers/v25/23-1547.html}
\BIBentrySTDinterwordspacing

\bibitem{deletang2024language}
\BIBentryALTinterwordspacing
G.~Deletang, A.~Ruoss, P.-A. Duquenne, E.~Catt, T.~Genewein, C.~Mattern,
  J.~Grau-Moya, L.~K. Wenliang, M.~Aitchison, L.~Orseau, M.~Hutter, and
  J.~Veness, ``Language modeling is compression,'' in \emph{The Twelfth
  International Conference on Learning Representations}, 2024. [Online].
  Available: \url{https://openreview.net/forum?id=jznbgiynus}
\BIBentrySTDinterwordspacing

\bibitem{huang2024compression}
\BIBentryALTinterwordspacing
Y.~Huang, J.~Zhang, Z.~Shan, and J.~He, ``Compression represents intelligence
  linearly,'' in \emph{First Conference on Language Modeling}, 2024. [Online].
  Available: \url{https://openreview.net/forum?id=SHMj84U5SH}
\BIBentrySTDinterwordspacing

\bibitem{li2025lossless}
Z.~Li, C.~Huang, X.~Wang, H.~Hu, C.~Wyeth, D.~Bu, Q.~Yu, W.~Gao, X.~Liu, and
  M.~Li, ``Lossless data compression by large models,'' \emph{Nature Machine
  Intelligence}, pp. 1--6, 2025.

\bibitem{wang2007logic}
P.~Wang, ``The logic of intelligence,'' in \emph{Artificial general
  intelligence}.\hskip 1em plus 0.5em minus 0.4em\relax Springer, 2007, pp.
  31--62.

\bibitem{feng2023data}
Z.~Feng, D.~Cai, Z.~Liu, J.~Shan, and W.~Wang, ``Data adaptive semantic
  communication systems for intelligent tasks and image transmission,'' in
  \emph{International Conference on AI-generated Content}.\hskip 1em plus 0.5em
  minus 0.4em\relax Springer, 2023, pp. 105--117.

\bibitem{joda2022internet}
R.~Joda, M.~Elsayed, H.~Abou-Zeid, R.~Atawia, A.~B. Sediq, G.~Boudreau,
  M.~Erol-Kantarci, and L.~Hanzo, ``The internet of senses: Building on
  semantic communications and edge intelligence,'' \emph{IEEE Network},
  vol.~37, no.~3, pp. 68--75, 2022.

\bibitem{dai2023toward}
J.~Dai, S.~Wang, K.~Yang, K.~Tan, X.~Qin, Z.~Si, K.~Niu, and P.~Zhang, ``Toward
  adaptive semantic communications: Efficient data transmission via online
  learned nonlinear transform source-channel coding,'' \emph{IEEE Journal on
  Selected Areas in Communications}, vol.~41, no.~8, pp. 2609--2627, 2023.

\bibitem{shannon1948mathematical}
C.~E. Shannon, ``A mathematical theory of communication,'' \emph{The Bell
  system technical journal}, vol.~27, no.~3, pp. 379--423, 1948.

\bibitem{weaver1949recent}
W.~Weaver, ``Recent contributions to the mathematical theory of
  communication,'' 1949.

\bibitem{carnap1952outline}
R.~Carnap and Y.~Bar-Hillel, ``An outline of a theory of semantic
  information,'' 1952.

\bibitem{bar1953semantic}
Y.~Bar-Hillel and R.~Carnap, ``Semantic information,'' \emph{The British
  journal for the philosophy of science}, vol.~4, no.~14, pp. 147--157, 1953.

\bibitem{bao2011towards}
J.~Bao, P.~Basu, M.~Dean, C.~Partridge, A.~Swami, W.~Leland, and J.~A. Hendler,
  ``Towards a theory of semantic communication,'' in \emph{2011 IEEE Network
  Science Workshop}.\hskip 1em plus 0.5em minus 0.4em\relax IEEE, 2011, pp.
  110--117.

\bibitem{de1972definition}
A.~De~Luca and S.~Termini, ``A definition of a nonprobabilistic entropy in the
  setting of fuzzy sets theory,'' \emph{INFORMATION AND CONTROL}, vol.~20, pp.
  301--312, 1972.

\bibitem{de1974entropy}
------, ``Entropy of l-fuzzy sets,'' \emph{Information and control}, vol.~24,
  no.~1, pp. 55--73, 1974.

\bibitem{seising200960}
R.~Seising, ``60 years ``a mathematical theory of communication''-towards a
  ``fuzzy information theory'','' in \emph{IFSA/EUSFLAT Conf.}, 2009, pp.
  1332--1337.

\bibitem{niu2024mathematical}
\BIBentryALTinterwordspacing
K.~Niu and P.~Zhang, ``A mathematical theory of semantic communication,''
  \emph{Journal on Communications}, vol.~45, no.~6, pp. 7--59, 2024. [Online].
  Available:
  \url{https://www.joconline.com.cn/en/article/doi/10.11959/j.issn.1000-436x.2024111/}
\BIBentrySTDinterwordspacing

\bibitem{niu2025mathematical}
------, \emph{The Mathematical Theory of Semantic Communication}.\hskip 1em
  plus 0.5em minus 0.4em\relax Springer Nature, 2025.

\bibitem{zhao2023survey}
W.~X. Zhao, K.~Zhou, J.~Li, T.~Tang, X.~Wang, Y.~Hou, Y.~Min, B.~Zhang,
  J.~Zhang, Z.~Dong \emph{et~al.}, ``A survey of large language models,''
  \emph{arXiv preprint arXiv:2303.18223}, 2023.

\bibitem{zhu2023minigpt}
D.~Zhu, J.~Chen, X.~Shen, X.~Li, and M.~Elhoseiny, ``Mini{GPT}-4: Enhancing
  vision-language understanding with advanced large language models,''
  \emph{arXiv preprint arXiv:2304.10592}, 2023.

\bibitem{xiao2022imitation}
Y.~Xiao, Z.~Sun, G.~Shi, and D.~Niyato, ``Imitation learning-based implicit
  semantic-aware communication networks: Multi-layer representation and
  collaborative reasoning,'' \emph{IEEE Journal on Selected Areas in
  Communications}, vol.~41, no.~3, pp. 639--658, 2022.

\bibitem{tian2024kg}
S.~Tian, Y.~Luo, T.~Xu, C.~Yuan, H.~Jiang, C.~Wei, and X.~Wang, ``{KG}-adapter:
  Enabling knowledge graph integration in large language models through
  parameter-efficient fine-tuning,'' in \emph{Findings of the Association for
  Computational Linguistics ACL 2024}, 2024, pp. 3813--3828.

\bibitem{zhang2022toward}
P.~Zhang, W.~Xu, H.~Gao, K.~Niu, X.~Xu, X.~Qin, C.~Yuan, Z.~Qin, H.~Zhao,
  J.~Wei \emph{et~al.}, ``Toward wisdom-evolutionary and primitive-concise
  {6G}: A new paradigm of semantic communication networks,''
  \emph{Engineering}, vol.~8, pp. 60--73, 2022.

\bibitem{zhang2024advances}
P.~Zhang, Y.~Liu, Y.~Song, and J.~Zhang, ``Advances and challenges in semantic
  communications: A systematic review,'' \emph{National Science Open}, vol.~3,
  no.~4, p. 20230029, 2024.

\bibitem{zheng2023semantic}
Y.~Zheng, F.~Wang, W.~Xu, M.~Pan, and P.~Zhang, ``Semantic communications with
  explicit semantic base for image transmission,'' in \emph{GLOBECOM 2023-2023
  IEEE Global Communications Conference}.\hskip 1em plus 0.5em minus
  0.4em\relax IEEE, 2023, pp. 4497--4502.

\bibitem{wang2024semantic}
F.~Wang, Y.~Zheng, W.~Xu, J.~Liang, and P.~Zhang, ``Semantic communications
  with explicit semantic bases: Model, architecture, and open problems,''
  \emph{arXiv preprint arXiv:2408.05596}, 2024.

\bibitem{liang2025synonymous}
\BIBentryALTinterwordspacing
Z.~Liang, K.~Niu, C.~Wang, J.~Xu, and P.~Zhang, ``Synonymous variational
  inference for perceptual image compression,'' in \emph{Proceedings of the
  42nd International Conference on Machine Learning (ICML)}, 2025. [Online].
  Available: \url{https://openreview.net/forum?id=ialr09SfeJ}
\BIBentrySTDinterwordspacing

\bibitem{blau2018perception}
Y.~Blau and T.~Michaeli, ``The perception-distortion tradeoff,'' in
  \emph{Proceedings of the IEEE conference on computer vision and pattern
  recognition}, 2018, pp. 6228--6237.

\bibitem{blau2019rethinking}
------, ``Rethinking lossy compression: The rate-distortion-perception
  tradeoff,'' in \emph{International Conference on Machine Learning}.\hskip 1em
  plus 0.5em minus 0.4em\relax PMLR, 2019, pp. 675--685.

\bibitem{vembu2002source}
S.~Vembu, S.~Verdu, and Y.~Steinberg, ``The source-channel separation theorem
  revisited,'' \emph{IEEE Transactions on Information Theory}, vol.~41, no.~1,
  pp. 44--54, 2002.

\bibitem{liang2024information}
Z.~Liang, K.~Niu, and P.~Zhang, ``Information conductivity: Universal
  performance measure for semantic communications,'' \emph{China
  Communications}, vol.~21, no.~7, pp. 17--36, 2024.

\bibitem{farsad2018deep}
N.~Farsad, M.~Rao, and A.~Goldsmith, ``Deep learning for joint source-channel
  coding of text,'' in \emph{2018 IEEE international conference on acoustics,
  speech and signal processing (ICASSP)}.\hskip 1em plus 0.5em minus
  0.4em\relax IEEE, 2018, pp. 2326--2330.

\bibitem{bourtsoulatze2019deep}
E.~Bourtsoulatze, D.~B. Kurka, and D.~G{\"u}nd{\"u}z, ``Deep joint
  source-channel coding for wireless image transmission,'' \emph{IEEE
  Transactions on Cognitive Communications and Networking}, vol.~5, no.~3, pp.
  567--579, 2019.

\bibitem{xie2021deep}
H.~Xie, Z.~Qin, G.~Y. Li, and B.-H. Juang, ``Deep learning enabled semantic
  communication systems,'' \emph{IEEE transactions on signal processing},
  vol.~69, pp. 2663--2675, 2021.

\bibitem{dai2022nonlinear}
J.~Dai, S.~Wang, K.~Tan, Z.~Si, X.~Qin, K.~Niu, and P.~Zhang, ``Nonlinear
  transform source-channel coding for semantic communications,'' \emph{IEEE
  Journal on Selected Areas in Communications}, vol.~40, no.~8, pp. 2300--2316,
  2022.

\bibitem{balle2018variational}
J.~Ball{\'e}, D.~Minnen, S.~Singh, S.~J. Hwang, and N.~Johnston, ``Variational
  image compression with a scale hyperprior,'' in \emph{International
  Conference on Learning Representations}, 2018.

\bibitem{balle2020nonlinear}
J.~Ball{\'e}, P.~A. Chou, D.~Minnen, S.~Singh, N.~Johnston, E.~Agustsson, S.~J.
  Hwang, and G.~Toderici, ``Nonlinear transform coding,'' \emph{IEEE Journal of
  Selected Topics in Signal Processing}, vol.~15, no.~2, pp. 339--353, 2020.

\bibitem{xiao2023wireless}
Z.~Xiao, S.~Yao, J.~Dai, S.~Wang, K.~Niu, and P.~Zhang, ``Wireless deep speech
  semantic transmission,'' in \emph{ICASSP 2023-2023 IEEE International
  Conference on Acoustics, Speech and Signal Processing (ICASSP)}.\hskip 1em
  plus 0.5em minus 0.4em\relax IEEE, 2023, pp. 1--5.

\bibitem{wang2022wireless}
S.~Wang, J.~Dai, Z.~Liang, K.~Niu, Z.~Si, C.~Dong, X.~Qin, and P.~Zhang,
  ``Wireless deep video semantic transmission,'' \emph{IEEE Journal on Selected
  Areas in Communications}, vol.~41, no.~1, pp. 214--229, 2022.

\bibitem{yue2023learned}
W.~Yue, J.~Dai, S.~Wang, Z.~Si, and K.~Niu, ``Learned source and channel coding
  for talking-head semantic transmission,'' in \emph{2023 IEEE Wireless
  Communications and Networking Conference (WCNC)}.\hskip 1em plus 0.5em minus
  0.4em\relax IEEE, 2023, pp. 1--6.

\bibitem{zhang2025toward}
G.~Zhang, K.~Zhou, Y.~Cai, Q.~Hu, and G.~Yu, ``Toward compatible semantic
  communication: A perspective on digital coding and modulation,'' \emph{IEEE
  Communications Magazine}, 2025.

\bibitem{bao2025sdac}
Z.~Bao, H.~Liange, X.~Liu, C.~Li, C.~Dong, X.~Xu, C.~Guo, H.~Chen, and
  P.~Zhang, ``s{DAC}—semantic digital analog converter for semantic
  communications,'' \emph{IEEE Transactions on Communications}, 2025.

\bibitem{bo2024joint}
Y.~Bo, Y.~Duan, S.~Shao, and M.~Tao, ``Joint coding-modulation for digital
  semantic communications via variational autoencoder,'' \emph{IEEE
  Transactions on Communications}, vol.~72, no.~9, pp. 5626--5640, 2024.

\bibitem{liu2024ofdm}
C.~Liu, C.~Guo, Y.~Yang, W.~Ni, and T.~Q. Quek, ``{OFDM}-based digital semantic
  communication with importance awareness,'' \emph{IEEE Transactions on
  Communications}, vol.~72, no.~10, pp. 6301--6315, 2024.

\bibitem{fu2023vector}
Q.~Fu, H.~Xie, Z.~Qin, G.~Slabaugh, and X.~Tao, ``Vector quantized semantic
  communication system,'' \emph{IEEE Wireless Communications Letters}, vol.~12,
  no.~6, pp. 982--986, 2023.

\bibitem{park2024joint}
J.~Park, Y.~Oh, S.~Kim, and Y.-S. Jeon, ``Joint source-channel coding for
  channel-adaptive digital semantic communications,'' \emph{IEEE Transactions
  on Cognitive Communications and Networking}, vol.~11, no.~1, pp. 75--89,
  2024.

\bibitem{zhang2024analog}
G.~Zhang, P.~Yang, Y.~Cai, Q.~Hu, and G.~Yu, ``From analog to digital:
  Multi-order digital joint coding-modulation for semantic communication,''
  \emph{IEEE Transactions on Communications}, 2024.

\bibitem{takeda2020understanding}
K.~Takeda, H.~Xu, T.~Kim, K.~Schober, and X.~Lin, ``Understanding the heart of
  the {5G} air interface: An overview of physical downlink control channel for
  {5G} new radio,'' \emph{IEEE Communications Standards Magazine}, vol.~4,
  no.~3, pp. 22--29, 2020.

\bibitem{xu2021wireless}
J.~Xu, B.~Ai, W.~Chen, A.~Yang, P.~Sun, and M.~Rodrigues, ``Wireless image
  transmission using deep source channel coding with attention modules,''
  \emph{IEEE Transactions on Circuits and Systems for Video Technology},
  vol.~32, no.~4, pp. 2315--2328, 2021.

\bibitem{xu2023deep}
J.~Xu, T.-Y. Tung, B.~Ai, W.~Chen, Y.~Sun, and D.~G{\"u}nd{\"u}z, ``Deep joint
  source-channel coding for semantic communications,'' \emph{IEEE
  communications Magazine}, vol.~61, no.~11, pp. 42--48, 2023.

\bibitem{jia2022rate}
C.~Jia, Z.~Ge, S.~Wang, S.~Ma, and W.~Gao, ``Rate distortion characteristic
  modeling for neural image compression,'' in \emph{2022 Data Compression
  Conference (DCC)}.\hskip 1em plus 0.5em minus 0.4em\relax IEEE, 2022, pp.
  202--211.

\bibitem{yang2023witt}
K.~Yang, S.~Wang, J.~Dai, K.~Tan, K.~Niu, and P.~Zhang, ``{WITT}: A wireless
  image transmission transformer for semantic communications,'' in \emph{ICASSP
  2023-2023 IEEE International Conference on Acoustics, Speech and Signal
  Processing (ICASSP)}.\hskip 1em plus 0.5em minus 0.4em\relax IEEE, 2023, pp.
  1--5.

\bibitem{yang2024swinjscc}
K.~Yang, S.~Wang, J.~Dai, X.~Qin, K.~Niu, and P.~Zhang, ``Swin{JSCC}: Taming
  swin transformer for deep joint source-channel coding,'' \emph{IEEE
  Transactions on Cognitive Communications and Networking}, 2024.

\bibitem{wu2024deep}
H.~Wu, Y.~Shao, C.~Bian, K.~Mikolajczyk, and D.~G{\"u}nd{\"u}z, ``Deep joint
  source-channel coding for adaptive image transmission over {MIMO} channels,''
  \emph{IEEE Transactions on Wireless Communications}, 2024.

\bibitem{zhang2024scan}
G.~Zhang, Q.~Hu, Y.~Cai, and G.~Yu, ``{SCAN}: Semantic communication with
  adaptive channel feedback,'' \emph{IEEE Transactions on Cognitive
  Communications and Networking}, vol.~10, no.~5, pp. 1759--1773, 2024.

\bibitem{xie2024robust}
B.~Xie, Y.~Wu, Y.~Shi, W.~Zhang, S.~Cui, and M.~Debbah, ``Robust image semantic
  coding with learnable {CSI} fusion masking over {MIMO} fading channels,''
  \emph{IEEE Transactions on Wireless Communications}, vol.~23, no.~10, pp.
  14\,155--14\,170, 2024.

\bibitem{lin2023channel}
L.~Lin, W.~Xu, F.~Wang, Y.~Zhang, W.~Zhang, and P.~Zhang,
  ``Channel-transferable semantic communications for multi-user {OFDM}-{NOMA}
  systems,'' \emph{IEEE Wireless Communications Letters}, vol.~13, no.~3, pp.
  721--725, 2023.

\bibitem{yao2023learned}
S.~Yao, S.~Wang, J.~Dai, and K.~Niu, ``Learned image transmission over {MIMO}
  fading channels,'' in \emph{2023 IEEE 34th Annual International Symposium on
  Personal, Indoor and Mobile Radio Communications (PIMRC)}.\hskip 1em plus
  0.5em minus 0.4em\relax IEEE, 2023, pp. 1--6.

\bibitem{guo2022overview}
J.~Guo, C.-K. Wen, S.~Jin, and G.~Y. Li, ``Overview of deep learning-based
  {CSI} feedback in massive mimo systems,'' \emph{IEEE Transactions on
  Communications}, vol.~70, no.~12, pp. 8017--8045, 2022.

\bibitem{wen2018deep}
C.-K. Wen, W.-T. Shih, and S.~Jin, ``Deep learning for massive {MIMO} {CSI}
  feedback,'' \emph{IEEE Wireless Communications Letters}, vol.~7, no.~5, pp.
  748--751, 2018.

\bibitem{yang2019deep}
Q.~Yang, M.~B. Mashhadi, and D.~G{\"u}nd{\"u}z, ``Deep convolutional
  compression for massive {MIMO} {CSI} feedback,'' in \emph{2019 IEEE 29th
  international workshop on machine learning for signal processing
  (MLSP)}.\hskip 1em plus 0.5em minus 0.4em\relax IEEE, 2019, pp. 1--6.

\bibitem{cui2022transnet}
Y.~Cui, A.~Guo, and C.~Song, ``Transnet: Full attention network for {CSI}
  feedback in {FDD} massive {MIMO} system,'' \emph{IEEE Wireless Communications
  Letters}, vol.~11, no.~5, pp. 903--907, 2022.

\bibitem{yin2022deep}
Z.~Yin, W.~Xu, R.~Xie, S.~Zhang, D.~W.~K. Ng, and X.~You, ``Deep {CSI}
  compression for massive {MIMO}: A self-information model-driven neural
  network,'' \emph{IEEE Transactions on Wireless Communications}, vol.~21,
  no.~10, pp. 8872--8886, 2022.

\bibitem{shin2024vector}
J.~Shin, Y.~Kang, and Y.-S. Jeon, ``Vector quantization for deep-learning-based
  {CSI} feedback in massive {MIMO} systems,'' \emph{IEEE Wireless
  Communications Letters}, vol.~13, no.~9, pp. 2382--2386, 2024.

\bibitem{lee2024learning}
J.~Lee, J.~Kim, and Y.~Choi, ``Learning variable-rate {CSI} compression with
  multi-stage vector quantization,'' in \emph{ICC 2024-IEEE International
  Conference on Communications}.\hskip 1em plus 0.5em minus 0.4em\relax IEEE,
  2024, pp. 5123--5128.

\bibitem{shin2025entropy}
J.~Shin, J.~Park, and Y.-S. Jeon, ``Entropy-constrained {VQ}-{VAE} for
  deep-learning-based {CSI} feedback,'' \emph{IEEE Transactions on Vehicular
  Technology}, 2025.

\bibitem{xu2022deep}
J.~Xu, B.~Ai, N.~Wang, and W.~Chen, ``Deep joint source-channel coding for
  {CSI} feedback: An end-to-end approach,'' \emph{IEEE Journal on Selected
  Areas in Communications}, vol.~41, no.~1, pp. 260--273, 2022.

\bibitem{zheng2025semantic}
Z.~Zheng, H.~Liang, Y.~Liu, C.~Dong, X.~Xu, and L.~Li, ``Semantic diversity for
  massive {MIMO} {CSI} feedback,'' in \emph{2025 10th International Conference
  on Computer and Communication System (ICCCS)}.\hskip 1em plus 0.5em minus
  0.4em\relax IEEE, 2025, pp. 831--837.

\bibitem{zhao2022joint}
L.~Zhao, H.~Xu, Z.~Wang, X.~Chen, and A.~Zhou, ``Joint channel estimation and
  feedback for mm-wave system using federated learning,'' \emph{IEEE
  communications letters}, vol.~26, no.~8, pp. 1819--1823, 2022.

\bibitem{guo2024deep}
J.~Guo, T.~Chen, S.~Jin, G.~Y. Li, X.~Wang, and X.~Hou, ``Deep learning for
  joint channel estimation and feedback in massive {MIMO} systems,''
  \emph{Digital Communications and Networks}, vol.~10, no.~1, pp. 83--93, 2024.

\bibitem{feng2024deep}
H.~Feng, Y.~Xu, and Y.~Zhao, ``Deep learning-based joint channel estimation and
  {CSI} feedback for {RIS}-assisted communications,'' \emph{IEEE Communications
  Letters}, vol.~28, no.~8, pp. 1860--1864, 2024.

\bibitem{guo2020deep}
J.~Guo, C.-K. Wen, and S.~Jin, ``Deep learning-based {CSI} feedback for
  beamforming in single-and multi-cell massive {MIMO} systems,'' \emph{IEEE
  Journal on Selected Areas in Communications}, vol.~39, no.~7, pp. 1872--1884,
  2020.

\bibitem{guo2024deep_multiuser}
Y.~Guo, W.~Chen, J.~Xu, L.~Li, and B.~Ai, ``Deep joint {CSI} feedback and
  multiuser precoding for {MIMO} {OFDM} systems,'' \emph{IEEE Transactions on
  Vehicular Technology}, 2024.

\bibitem{guo2025deep}
Y.~Guo, W.~Chen, B.~Ai, and L.~Li, ``Deep joint {CSI}
  estimation-feedback-precoding for {MU}-{MIMO} {OFDM} systems,'' \emph{arXiv
  preprint arXiv:2503.04157}, 2025.

\bibitem{wang2024digital}
H.~Wang, J.~Zhang, G.~Nie, L.~Yu, Z.~Yuan, T.~Li, J.~Wang, and G.~Liu,
  ``Digital twin channel for {6G}: Concepts, architectures and potential
  applications,'' \emph{IEEE Communications Magazine}, 2024.

\bibitem{wang2025radio}
J.~Wang, J.~Zhang, Y.~Zhang, Y.~Sun, G.~Nie, L.~Shi, P.~Zhang, and G.~Liu,
  ``Radio environment knowledge pool for {6G} digital twin channel,''
  \emph{IEEE Communications Magazine}, 2025.

\bibitem{shah2021survey}
A.~S. Shah, A.~N. Qasim, M.~A. Karabulut, H.~Ilhan, and M.~B. Islam, ``Survey
  and performance evaluation of multiple access schemes for next-generation
  wireless communication systems,'' \emph{IEEe Access}, vol.~9, pp.
  113\,428--113\,442, 2021.

\bibitem{yonis2012lte}
A.~Yonis, M.~F.~L. Abdullah, and M.~Ghanim, ``{LTE}-{FDD} and {LTE}-{TDD} for
  cellular communications,'' \emph{Proceeding, Progress in}, 2012.

\bibitem{el2011network}
A.~El~Gamal and Y.-H. Kim, \emph{Network information theory}.\hskip 1em plus
  0.5em minus 0.4em\relax Cambridge university press, 2011.

\bibitem{ding2017survey}
Z.~Ding, X.~Lei, G.~K. Karagiannidis, R.~Schober, J.~Yuan, and V.~K. Bhargava,
  ``A survey on non-orthogonal multiple access for {5G} networks: Research
  challenges and future trends,'' \emph{IEEE Journal on Selected Areas in
  Communications}, vol.~35, no.~10, pp. 2181--2195, 2017.

\bibitem{uyoata2021relaying}
U.~Uyoata, J.~Mwangama, and R.~Adeogun, ``Relaying in the internet of things
  ({IoT}): A survey,'' \emph{Ieee Access}, vol.~9, pp. 132\,675--132\,704,
  2021.

\bibitem{siddiqui2021interference}
M.~U.~A. Siddiqui, F.~Qamar, F.~Ahmed, Q.~N. Nguyen, and R.~Hassan,
  ``Interference management in {5G} and beyond network: Requirements,
  challenges and future directions,'' \emph{IEEE Access}, vol.~9, pp.
  68\,932--68\,965, 2021.

\bibitem{kolodziej2019band}
K.~E. Kolodziej, B.~T. Perry, and J.~S. Herd, ``In-band full-duplex technology:
  Techniques and systems survey,'' \emph{IEEE Transactions on Microwave Theory
  and Techniques}, vol.~67, no.~7, pp. 3025--3041, 2019.

\bibitem{zhang2023model}
P.~Zhang, X.~Xu, C.~Dong, K.~Niu, H.~Liang, Z.~Liang, X.~Qin, M.~Sun, H.~Chen,
  N.~Ma \emph{et~al.}, ``Model division multiple access for semantic
  communications,'' \emph{Frontiers of Information Technology \& Electronic
  Engineering}, vol.~24, no.~6, pp. 801--812, 2023.

\bibitem{niu2024semantics}
K.~Niu, Z.~Liang, C.~Dong, J.~Dai, Z.~Si, and P.~Zhang, ``Semantics-division
  duplexing: A novel full-duplex paradigm,'' \emph{IEEE Wireless
  Communications}, vol.~31, no.~4, pp. 307--314, 2024.

\bibitem{cover1999elements}
T.~M. Cover, \emph{Elements of information theory}.\hskip 1em plus 0.5em minus
  0.4em\relax John Wiley \& Sons, 1999.

\bibitem{zhang2023deepma}
W.~Zhang, K.~Bai, S.~Zeadally, H.~Zhang, H.~Shao, H.~Ma, and V.~C. Leung,
  ``Deep{MA}: End-to-end deep multiple access for wireless image transmission
  in semantic communication,'' \emph{IEEE transactions on cognitive
  communications and networking}, vol.~10, no.~2, pp. 387--402, 2023.

\bibitem{li2023non}
W.~Li, H.~Liang, C.~Dong, X.~Xu, P.~Zhang, and K.~Liu, ``Non-orthogonal
  multiple access enhanced multi-user semantic communication,'' \emph{IEEE
  Transactions on Cognitive Communications and Networking}, vol.~9, no.~6, pp.
  1438--1453, 2023.

\bibitem{mu2023exploiting}
X.~Mu and Y.~Liu, ``Exploiting semantic communication for non-orthogonal
  multiple access,'' \emph{IEEE Journal on Selected Areas in Communications},
  vol.~41, no.~8, pp. 2563--2576, 2023.

\bibitem{liang2024orthogonal}
H.~Liang, K.~Liu, X.~Liu, H.~Jiang, C.~Dong, X.~Xu, K.~Niu, and P.~Zhang,
  ``Orthogonal model division multiple access,'' \emph{IEEE Transactions on
  Wireless Communications}, vol.~23, no.~9, pp. 11\,693--11\,707, 2024.

\bibitem{luo2022autoencoder}
X.~Luo, B.~Yin, Z.~Chen, B.~Xia, and J.~Wang, ``Autoencoder-based semantic
  communication systems with relay channels,'' in \emph{2022 IEEE International
  Conference on Communications Workshops (ICC Workshops)}.\hskip 1em plus 0.5em
  minus 0.4em\relax IEEE, 2022, pp. 711--716.

\bibitem{lin2024semantic}
W.~Lin, Y.~Yan, L.~Li, Z.~Han, and T.~Matsumoto, ``Semantic-forward relaying: A
  novel framework toward {6G} cooperative communications,'' \emph{IEEE
  Communications Letters}, vol.~28, no.~3, pp. 518--522, 2024.

\bibitem{hu2024semantic}
Z.~Hu, C.~You, T.~Liu, D.~Wen, Y.~Hu, Y.~Cui, Y.~Gong, and K.~Huang, ``Semantic
  communication meets edge intelligence: Semantic-relay-aided text
  transmissions,'' \emph{IEEE Internet of Things Journal}, 2024.

\bibitem{wu2025icdm}
T.~Wu, Z.~Chen, D.~He, F.~Yang, M.~Tao, X.~Xu, W.~Zhang, and P.~Zhang,
  ``{ICDM}: Interference cancellation diffusion models for wireless semantic
  communications,'' \emph{arXiv preprint arXiv:2505.19983}, 2025.

\bibitem{ma2024semantic}
S.~Ma, C.~Zhang, B.~Shen, Y.~Wu, H.~Li, S.~Li, G.~Shi, and N.~Al-Dhahir,
  ``Semantic feature division multiple access for multi-user digital
  interference networks,'' \emph{IEEE Transactions on Wireless Communications},
  2024.

\bibitem{zhang2024interference}
Y.~Zhang, R.~Zhong, Y.~Liu, W.~Xu, and P.~Zhang, ``Interference suppressed
  {NOMA} for semantic-aware communication networks,'' \emph{IEEE Transactions
  on Wireless Communications}, vol.~23, no.~8, pp. 10\,383--10\,397, 2024.

\bibitem{mazandarani2024semantic}
H.~Mazandarani, M.~Shokrnezhad, and T.~Taleb, ``A semantic-aware multiple
  access scheme for distributed, dynamic 6g-based applications,'' in \emph{2024
  IEEE Wireless Communications and Networking Conference (WCNC)}.\hskip 1em
  plus 0.5em minus 0.4em\relax IEEE, 2024, pp. 1--6.

\bibitem{guo2024distributed}
J.~Guo, H.~Chen, B.~Song, Y.~Chi, C.~Yuen, F.~R. Yu, G.~Y. Li, and D.~Niyato,
  ``Distributed task-oriented communication networks with multimodal semantic
  relay and edge intelligence,'' \emph{IEEE Communications Magazine}, vol.~62,
  no.~6, pp. 82--89, 2024.

\bibitem{du2023ai}
H.~Du, J.~Wang, D.~Niyato, J.~Kang, Z.~Xiong, and D.~I. Kim, ``{AI}-generated
  incentive mechanism and full-duplex semantic communications for information
  sharing,'' \emph{IEEE Journal on Selected Areas in Communications}, vol.~41,
  no.~9, pp. 2981--2997, 2023.

\bibitem{bossen2010common}
F.~Bossen, ``Common test conditions and software reference configurations,'' in
  \emph{3rd. JCT-VC Meeting, Guangzhou, CN, October 2010}, 2010.

\bibitem{qi2024wafi}
X.~Qi, N.~Ma, Z.~Bao, Y.~Liu, C.~Dong, and X.~Xu, ``{WAFI-VSC}: Wireless
  adaptive frame interpolation video semantic communication,'' in \emph{2024
  16th International Conference on Wireless Communications and Signal
  Processing (WCSP)}.\hskip 1em plus 0.5em minus 0.4em\relax IEEE, 2024, pp.
  1503--1508.

\bibitem{kurka2020deepjscc}
D.~B. Kurka and D.~G{\"u}nd{\"u}z, ``Deep{JSCC}-f: Deep joint source-channel
  coding of images with feedback,'' \emph{IEEE journal on selected areas in
  information theory}, vol.~1, no.~1, pp. 178--193, 2020.

\bibitem{wang2003multiscale}
Z.~Wang, E.~P. Simoncelli, and A.~C. Bovik, ``Multiscale structural similarity
  for image quality assessment,'' in \emph{The thrity-seventh asilomar
  conference on signals, systems \& computers, 2003}, vol.~2.\hskip 1em plus
  0.5em minus 0.4em\relax Ieee, 2003, pp. 1398--1402.

\bibitem{wiegand2003overview}
T.~Wiegand, G.~J. Sullivan, G.~Bjontegaard, and A.~Luthra, ``Overview of the
  {H}. 264/{AVC} video coding standard,'' \emph{IEEE Transactions on circuits
  and systems for video technology}, vol.~13, no.~7, pp. 560--576, 2003.

\bibitem{sullivan2012overview}
G.~J. Sullivan, J.-R. Ohm, W.-J. Han, and T.~Wiegand, ``Overview of the high
  efficiency video coding ({HEVC}) standard,'' \emph{IEEE Transactions on
  circuits and systems for video technology}, vol.~22, no.~12, pp. 1649--1668,
  2012.

\bibitem{gallager2003low}
R.~Gallager, ``Low-density parity-check codes,'' \emph{IRE Transactions on
  information theory}, vol.~8, no.~1, pp. 21--28, 2003.

\end{thebibliography}

\AtEndDocument{\par\leavevmode}
\bibliographystyle{IEEEtran}

\end{document}